\documentclass[manuscript]{acmart}
\AtBeginDocument{%
  }
\usepackage[utf8]{inputenc}
\usepackage{array}

\usepackage{amssymb}   
\usepackage{subfig}
\usepackage{enumitem}

\usepackage{tabularx}
\usepackage{array}
\usepackage{longtable}
\usepackage{booktabs}
\usepackage{comment}
\newcolumntype{L}{>{\raggedright\arraybackslash}X}

\setcopyright{acmlicensed}
\copyrightyear{2018}
\acmYear{2018}
\acmDOI{XXXXXXX.XXXXXXX}
\acmConference[Conference acronym 'XX]{Make sure to enter the correct
  conference title from your rights confirmation email}{June 03--05,
  2018}{Woodstock, NY}
\acmISBN{978-1-4503-XXXX-X/2018/06}

\begin{document}

\title{\edit{Allowing} Generative AI in CS1 with Code Review Interviews and a Flipped Classroom}


\author{Peter Fowles}
\affiliation{%
  \institution{Utah State University}
  \city{Logan}
  \state{Utah}
  \country{USA}}
\email{peter.fowles@usu.edu}

\author{Erik Falor}
\affiliation{%
  \institution{Utah State University}
  \city{Logan}
  \state{Utah}
  \country{USA}}
\email{erik.falor@usu.edu}

\author{Sulove Bhattarai}
\affiliation{%
  \institution{Utah State University}
  \city{Logan}
  \state{Utah}
  \country{USA}}
\email{sulove.bhattarai@usu.edu}

\author{John Edwards}
\affiliation{%
  \institution{Utah State University}
  \city{Logan}
  \state{Utah}
  \country{USA}}
\email{john.edwards@usu.edu}

\author{Seth Poulsen}
\affiliation{%
  \institution{Utah State University}
  \city{Logan}
  \state{Utah}
  \country{USA}}
\email{seth.poulsen@usu.edu}

\renewcommand{\shortauthors}{Fowles et al.}

\begin{abstract}

\textbf{Background and Context:} The rapid surge in Generative Artificial Intelligence (GenAI) technology has made Large Language Models (LLMs) more accessible and accurate than ever before. This surge in LLM usage has raised many significant concerns for computing educators and researchers. Notable among these issues are the capability of LLMs to generate working code in moments, bypassing the effort and learning processes needed to develop an understanding of concepts and metacognitive strategies essential for success in computer science. \\
\textbf{Objectives:} In this paper, we contribute a unique approach to assessing and building up student understanding through weekly oral code review assessments conducted by trained teaching assistants for an introductory computer science course at a university in the United States. These formative assessments are designed to incentivize students to make an active attempt at understanding their submitted code, regardless of whether or not the code was generated by AI tools. In this course, a flipped classroom environment is also implemented to provide time for students to take time to learn concepts outside of class and provide time for students to schedule code review interviews. \\
\textbf{Method:} For this paper, we collected exam scores from the Fall 2023, Fall 2024, and Fall 2025 semesters. We provide quantitative analyses of these datasets to discover how student performance was impacted by the new \edit{course policies} introduced in Fall 2025. We provide a comparison of keystroke log data from coding assignments submitted in Fall 2024 and Fall 2025 to show how AI usage may have changed with new policies. We also surveyed Fall 2025 students to gauge their perceptions towards code review policies, as well as identify the various ways they used AI tools and prepared for code review sessions through the semester.  \\
\textbf{Findings:} Pairwise comparison of exam results reveals \edit{no statistically significant change} in average scores for Fall 2025 compared to previous semesters. Keystroke logs show a significant increase in characters pasted per total characters input into coding assignments in Fall 2025, contributing evidence towards higher AI usage. Survey results show overwhelmingly positive student sentiment towards code review policies at the end of Fall 2025, with nearly all negative feedback being addressable through better scheduling and more rigorous TA training. \\
\textbf{Implications:} Oral code review assessments alongside a flipped classroom environment appear to be effective at mitigating \edit{potential} harms of generative AI use while providing space for students to freely experiment with these tools. Our work suggests that students in Fall 2025 still show adequate understanding of material covered in written exams, despite increases in generative AI usage for coding assignments. 
\end{abstract}




\received{20 February 2007}
\received[revised]{12 March 2009}
\received[accepted]{5 June 2009}

\newcommand{\todo}[1]{\textcolor{red}{TODO: #1}}
\newcommand{\edit}[1]{#1}

\maketitle

\section{Introduction}
 
Generative AI (GenAI) tools are currently transforming how education works in ways that cannot be avoided. The capabilities of these tools to perform tasks previously held to be essential to computing education---such as code generation and comprehension---are improving at alarming rates~\citep{prather2023weird, becker2023programming}. With widespread availability, AI tools have great potential to benefit students when used responsibly~\cite{zou2026depend}, but could have lasting stifling effects on students' cognitive development when used inappropriately~\citep{tian2025learners}.

Educators have made various attempts at either resisting this new technology in an appeal to maintain more traditional pedagogical approaches, or embracing AI coding tools to prepare future generations of students for an AI-oriented world~\cite{lau2023ban}.

In this paper, we present a new approach to embracing AI tools in an introductory computer science course. First implemented in a Fall 2025 CS1 course with 96 students, this approach requires students to demonstrate their understanding through a mandatory 15-minute code review interview conducted by a trained teaching assistant (TA) after every weekly coding assignment submission. This CS1 Code Review course (abbreviated to CS1-CR) has no restrictions on student usage of AI tools for coding assignments, so long as they are able to demonstrate their understanding and justify their implementation decisions during the interviews. CS1-CR also implements a flipped classroom learning environment \edit{to make up for class time lost to code reviews and encourage students to engage with the course content.}

To evaluate the effects of CS1-CR's new policies, we collected and compared exam scores and keystroke logs from the CS1-CR course and two CS1 courses taught in previous years. Additionally, we distributed an end-of-semester survey to evaluate student perceptions and self-reported behaviors throughout the CS1-CR course. These methods aim to answer the following research questions:

    \textbf{RQ1:} How is student performance impacted by policies of CS1-CR?
    
    \textbf{RQ2:} How is student usage of generative AI tools impacted by policies of CS1-CR?
    
    \textbf{RQ3:} How do students study and prepare for code review interviews in CS1-CR?
    
    \textbf{RQ4:} What are the students' perceptions of \edit{code review interviews} in CS1-CR? 

Results from our study indicate that CS1-CR enhances student understanding of course content and improves their learning experience. This is indicated by a 2\% improvement in CS1-CR exam scores when compared to previous semesters, in addition to overwhelmingly positive feedback from students. These positive results stand alongside keystroke logging data showing that CS1-CR students tended to paste more code into each assignment compared to the Fall 2024 semester, indicating higher AI usage. Overall, we found that CS1-CR provided a positive environment for students to engage in deeper learning by explaining their understanding to a trained peer. Despite a class size of 96 students, CS1-CR was able to scale its work to effectively provide the space for each student to participate in the code review process. 

    
\section{Related Work}

\subsection{How do we characterize student use of generative AI tools?}

Emerging research has begun to characterize the ways in which students use generative AI tools while working on programming assignments~\citep{ghimire2024coding,adeeb2025novice,brender2024s}. \citet{brender2024s} classified student interactions with generative AI into debugging, practical development, and conceptual exploring. Students who used AI for conceptual exploration had stronger learning gains than those who simply used it to generate code.  
\citet{ghimire2024coding} also analyzed student interactions with an AI tool, and came up with similar categorizations.
\citet{adeeb2025novice} also found that students could engage in independent problem solving even when given access to ChatGPT for problem solving help.


\subsection{Student learning gains while using generative AI tools}
A number of studies have measured both student productivity and learning gains while using AI tools, finding mixed results. 
While some students struggle to even evaluate short snippets of code generated by AI, higher performing students can have their productivity greatly accelerated~\cite[]{vaithilingam2022expectation,prather2024widening}.
As generally expected, students who overly rely on AI tools to complete their work for them can have improved productivity gains in completing the work but with poor learning gains~\cite[]{jovst2024impact,brender2024s,shihab2025effects}. However, there is also evidence that some types of engagement with AI tools, such as conceptual exploration, can boost student productivity while still maintaining or even boosting learning gains~\cite[]{xue2024does,brender2024s}.
Most studies measuring learning gains have been limited to laboratory studies. As of yet, there have been no randomized controlled trials. 


\citet{vadaparty2025integrating,vadaparty2024cs1} introduced CS1-LLM, a way of integrating generative AI tools into a classroom in a way that centers students understanding and communicating code understanding, with the view that future programmers will primarily work together with AI tools to generate code rather than writing it from scratch. They showed that their students could perform equally well on benchmark code tracing and code explaining questions as students who took a more traditional CS1 course without AI tools integrated. We see the approach of \citet{vadaparty2025integrating,vadaparty2024cs1} as complementary to ours. While they changed their course's learning outcomes in light of generative AI tools, the course we present in this paper, CS1-CR, keeps the same learning goals as a traditional CS1 course, but uses code review interviews to ensure students still meet the learning goals while programming with AI tools. Both approaches could be a valid way forward in the age of AI developer tools.

\subsection{Developer, instructor, and student perspectives on AI tools in Software Engineering and CS education}
The task of adapting classroom environments to safely and effectively make room for the seemingly inescapable AI technology poses a heavy burden for educators and researchers alike. Concerns about usage of AI for cheating on coding assignments have lead instructors to take measures to ban AI for coursework, weigh exam scores more heavily, and educate students on the limitations of these tools. However, instructors are well-aware that these measures cannot be taken indefinitely. In plans for the future, some respond to these concerns by re-working their assignments and other materials to be as ``AI-proof'' as possible. Others opt to embrace AI tools, experimenting with new techniques to make AI usage a key part of their course structures~\citep{lau2023ban}.

Recent research that interviews and surveys both academics and software engineering professionals shows a high level of agreement that in the age of AI-assisted software development, there are new skills that developers need to learn. The core of these are the ability to effectively prompt generative AI systems to generate code in the desired way and the ability to evaluate the code that is generated~\citep{kam2025what,prather2025beyond}. Many students and teachers feel that AI should and will play a role in education in the future, but it is not clear yet what the exact role should be, and how it should be best integrated in ways that are conducive to student learning~\citep{ghimire2024from,zastudil2023generative}.

\subsection{Code Reviews for Educational Purposes}
In higher educational contexts, the concept of peer assessment has been a highly researched subject for several decades~\cite{topping1998peer}. Peer code reviews have been extensively researched~\cite{indriasari2020peer}, with studies showing valuable benefits such as learning alternative implementation approaches, getting helpful feedback sooner than graders could reasonably provide it, and discussing coding practices~\cite{hundhausen2009review}.
Peer code reviews have been introduced into the classroom in many different ways, from individual student critique~\cite{turner2018peer}, to collaborative group review of each others' code~\cite{rivera2020review}. Our work focuses on instructor and TA review of student code, rather than peer code review.

\subsection{Oral Exams}
\label{sec:oralexams}
In a 2005 article, \citet{gharibyan2005oral} discusses a number of advantages that oral exams hold over traditional written exams. For the instructor/grader, oral exams are generally easier for the instructor to prepare, and give students a chance to demonstrate their understanding and knowledge with more flexible evaluations. Some possible challenges posed by oral exams include additional stress for students with anxiety towards interpersonal confrontation, as well as bias and unfairness from the examiner.

Oral exams have been studied to some extent in other STEM fields such as biology~\cite{huxham2012oral}, thermodynamics~\cite{zhao2018oral}, and mathematics~\cite{iannone2012oral}. However, there have only been a select few documented attempts at implementing oral exams in computer science courses. \citet{ohmann2019oral} presents one such attempt in a 2019 introductory CS course. In 2025, \citet{ohmann2025oral} provided a follow-up study, further evaluating implementations of oral exams in introductory CS courses at two institutions. These reports show positive experiences for both students and instructors with an oral final exam, supported by qualitative analysis of student feedback and quantitative comparison of instructor time commitments when grading written vs. oral exams. The authors point out a need for further research into oral exams in undergraduate-level CS courses, emphasizing a current knowledge gap regarding how to scale administration of oral exams for larger class sizes. 

\citet{grunwald2015personalized} provide an implementation of formative oral assessments in CS1 and CS2 courses in an effort to allow students to consult peers and other outside resources. These interviews received positive feedback from students, with the main complaints being grader inconsistency and scheduling logistics. \edit{A similar implementation in a 2005 systems analysis and design course had student feedback indicating the same benefits and potential downsides~\cite{dick2005interviews}.} CS1-CR adopts a number of common traits in its grading policy, though the intention is directed towards usage of AI tools, which were not nearly as relevant or useful at the time of this paper's publication (2015).

\subsection{\edit{Teaching Assistant Roles, Responsibilities, and Experiences}}
\label{sec:related_work_tas}
\edit{
Teaching Assistants (TAs) are individuals hired to assist instructors in their various responsibilities in teaching, grading, and other administrative tasks over a course. In undergraduate introductory computer science courses, TAs are generally either undergraduate or graduate students. These roles are generally done for pay via hourly time logging or stipends, and/or additional credit towards their degree.
}

\edit{
Prior work indicates that the most common responsibilities of undergraduate student TAs (UTAs) are providing assistance to students on coding assignments, reinforcing concepts taught in class for smaller sections, and grading student assignments, exams, and other work~\cite{mirza2019uta}. A 2025 survey of graduate student TAs (GTAs) conducted by \citet{mcleod2025gta} showed that GTAs generally have similar responsibilities to those described in other work pertaining to UTAs, albeit with a seemingly higher emphasis on leading lab sessions for subgroups of larger courses. This survey also indicates that most GTAs spend 11-20 hours per week on these responsibilities. 
}

\edit{
In computer science, research on TAs has mostly focused on the types of help students request in office hours or TA labs~\cite{ren2019what,ko2025relationships,doebling2021patterns} or TA management and training~\cite{mcdonald2023managing,zaman2025graduate}.
Beyond computer science, research on TAs has covered topics such as TA training, TA self-efficacy, and concerns that TAs may have with their duties~\cite{cho2011concerns,chiu2019study,elena2025perspectives}.
Existing research has shown that TAs can be apprehensive about some of their duties, especially when not given adequate training or experience~\cite{cho2011concerns,chiu2019study}.
On the other hand,
TAs view more direct student interaction positively~\cite{elena2025perspectives}, and having them do code reviews provides more opportunity for that, at a time when less students are attending office hours due to use of AI tools~\cite{mit2026aicommittee}. 
}

\edit{
There is a notable gap in the literature on TAs conducting oral examinations: to the authors' knowledge, no study has examined the expeirence of TAs conducting oral exams. Conducting oral exams is not mentioned at all in the existing TA self-efficacy scale~\cite{prieto1999effects,chiu2019study},  while it does cover many other common TA duties such as explaining concepts to students, running lab sections, or giving students feedback on \emph{written} exams. 
Therefore, from what we know in existing research, there are aspects of conducting code reviews which TAs may either views positively (increased student interaction) or negatively (apprehension toward new duties).
While this paper focuses on the student experience in a course that uses oral exams to evaluate students' completion of their programming assignments, our future work will also address understanding the TA perspective on administering the oral exams.
}


\subsection{Flipped Classroom Pedagogy}

In a flipped classroom environment, students are expected to study the required lecture materials before each class and participate in learning activities during class. This learning model is designed to allow students to review course materials at their own convenience and improve student engagement and learning by providing supplemental activities with interactive, personalized support from instructors and peers. This model also has the instructor prepare and distribute lecture materials before the course even begins, allowing the instructor to be free and present to monitor the learning of individual students during interactive learning activities~\cite{sams2012flip, sarawagi2014flipped}. Meta-analyses of flipped classrooms across many subject domains~\cite{cheng2018flipped, abidin2024flipped} show that flipped classrooms generally have small positive effects on learning, though the effects can vary across fields. In CS courses, usage of flipped classroom pedagogy has shown positive student sentiment and benefits to learning outcomes~\cite{chang2018flipped, durak2019flipped, mohamed2020flipped, alhazbi2016flipped}.

The instructor of CS1-CR chose to cancel class one out of three days each week to give the students time to complete their code review interviews, necessitating some change in instruction to cover the material in less class time. The instructor chose to address this issue by using a flipped classroom model, where students would watch lecture videos outside of class and then perform supplemental learning activities in the remaining two days of class each week.


\section{Course Context} \label{sec:context}
In this section, we describe the changes made in CS1-CR's structure and policies, contrasting them with those from previous iterations of CS1 courses. In addition to giving further context for our study, this section, alongside the code review interview protocol provided in Appendix~\ref{sec:protocol}, should supply instructors and researchers with sufficient resources to replicate or modify elements of CS1-CR in their own courses. 

\subsection{Course Structure}


The courses compared in this study were sections of a CS1 course taught over the Fall 2023 (Fa23), Fall 2024 (Fa24), and Fall 2025 (CS1-CR) semesters. Through these three semesters, the courses all shared the same instructor, course materials, quizzes, and exams. Coding assignments were generally the same but with a few modifications over the years, detailed in Table~\ref{tab: assignments used}. Only assignments that remained completely identical are used for the keystroke analysis. All three courses also shared the same overall weight distributions between assignments, quizzes, exams, and participation. Fa23 and Fa24 both followed a traditional lecture structure. CS1-CR deviated from this to implement a flipped classroom learning environment instead, largely to make up for the lost class time due to one out of three class days each week begin devoted to code review interviews. Students were provided with lecture videos to watch before class, with class time devoted to working through supplemental coding activities alongside peers under instructor supervision.

\begin{table}[htbp]
    \centering
    \caption{Programming assignments used in different semesters of the course. Over the three years, assignments remained the same with some exceptions: A3b (2024) was replaced with A3b' (CS1-CR); A4 (2024) was renamed A4b (CS1-CR). The content of A6 (2023) and A6b (2024, 2025) were the same. Keystroke analysis was only for Fall 2024 and CS1-CR as keystroke logs were not collected in Fall 2023 (see Section~\ref{sec:keystroke methods})} 
    \begin{tabular}{c|c|c|c|c|c|c|c|c|c|c|c|c|c|c|c} \hline
    Semester & \multicolumn{15}{c}{Assignments} \\ 
    \hline
    Fall 2023 & A1a & A1b & A2a & A2b & A3a & A3b &  &  & A4 & A5a &  & A5b &  & A6 & A7 \\
    Fall 2024 & A1a & A1b & A2a & A2b & A3a & A3b &  &  & A4 & A5a &  & A5b &  & A6b & A7 \\
    CS1-CR & A1a & A1b & A2a & A2b & A3a &  & A3b' & A4a & A4b &  & A5a' &  & A5b' & A6b & A7 
    \end{tabular}
    \label{tab: assignments used}
\end{table}

Fa23 and Fa24 both required students to complete all coding assignments entirely on their own, prohibiting any usage of GenAI tools for any assistance. CS1-CR saw this policy changed completely---rather than prohibiting or discouraging these tools, students were encouraged to embrace and experiment with GenAI in their coding assignments. To combat concerns about reliance on these tools stifling student learning and growth, students were also required to complete an oral, in-person code review (CR) interview session with a TA or instructor within 48 hours after each submission deadline. Students were able to schedule these sessions in advance or join a queue through the university's computer science tutoring center app. All sessions were held in the tutoring center, with the exception of those conducted by the course instructor or any that needed to be held over Zoom due to extenuating circumstances. These sessions would have the interviewer ask a set of targeted questions to evaluate how well the student understood the contents of their program and why certain decisions were made in their implementation. The interviewer would then have the student demonstrate their program's functionality by running a series of given integration test cases. The code review interview protocol is  provided in Appendix~\ref{sec:protocol}. 

\edit{Because code review interviews required an additional weekly time commitment for all students in CS1-CR, the instructor opted to switch to a flipped classroom environment. This new structure had two class sessions per week where students were given learning activities in class, while learning the actual course materials from recorded lectures outside of class. Previous semesters had three class sessions per week which followed a traditional lecture structure instead. The flipped classroom in CS1-CR was implemented primarily to give students extra time each week for scheduling their code review interviews. Further discussion of how this structure impacts the interpretation of our results, and our suggestions for future work, are provided in Sections~\ref{sec:limitations} and~\ref{sec:future_work}.
}

While the assignment contents and requirements were mostly identical across the years, the way assignments were graded was completely changed in CS1-CR. Fa23 and Fa24 had students' overall score on an assignment 100\% reflected by their rubric score (code quality, program behavior, adherence to assignment description, etc.). However, CS1-CR, while continuing to adhere to the same rubric for each coding assignment, only counted the rubric score for 30\% of each student's overall grade on the assignment, with the remaining 70\% coming from the CR sessions. Students were responsible for scheduling and completing a CR session within 48 hours of the assignment deadline. Failing to do so would result in an immediate zero on the \emph{entire} assignment grade. The instructor accommodated for this strict deadline by opting to have no class on Mondays, providing an extra hour every week for students to schedule sessions. 

Appendix~\ref{sec:protocol} provides a template example of the structured protocol given to TAs to guide their code review interviews.
Although the coding assignments across the three semesters all shared the same tasks and starter code, the tasks were split slightly differently across each semester. Table~\ref{tab: assignments used} shows each assignment used through all the semesters.

\subsection{TA training}

Before the first assignment deadline of the semester, the 11 TAs were trained by the course's instructor on how to properly conduct the interviews. This training session included a briefing, followed by a series of mock interviews to practice following the protocol. Alongside each assignment, the instructor also released a clearly structured interview protocol to ensure that the TAs were effectively evaluating the assignment's specific learning goals, modeled after the template shown in Appendix~\ref{sec:protocol}. 

\edit{
At the university where this was implemented, there was already an organized CS tutoring center that provided the staff and infrastructure needed for CS1-CR to run relatively smoothly. This also provides a foundation for further scaling by hiring more TAs and tutors and improving training and on-boarding procedures.
}

\section{Methods}

\subsection{Data Collection}
In our study, we use data from the Fa23, Fa24, and CS1-CR CS1 courses. \edit{Our reason for selecting these specific sections are that they each were taught by the same instructor and followed the same course outline, with identical exams and coding assignments. The only differences across the three sections are the usage of code review interviews and a flipped classroom environment in CS1-CR.} The following sections describe the various forms of data collected in greater detail. All data was scrubbed of any potentially identifiable information by the institution's approved student data handlers prior to them delivering it to the research team. This procedure was approved by the university's ethical review board under IRB \#xxxxx. 

\subsubsection{Test Scores} \label{sec:scores}
Because the AI policy was different in CS1-CR compared to Fa23 and Fa24, comparing grades on coding assignments would not be an effective way to evaluate how student performance was affected by the new AI and code review policies. Therefore, we determined that the most effective way to measure this effect would be to compare scores from the two proctored, auto-graded multiple choice exams held over each of the three semesters.

The set of questions, student answers, and an answer key were obtained for each of the two exams given in Fa23, Fa24, and CS1-CR. The questions on each of these exams assessed topics such as Python syntax, data types, arithmetic, logic, and core programming concepts, as well as skills such as tracing and reverse tracing through code snippets and given possible outputs. These were given in the form of multiple-choice and drag-and-drop style problems on a digital interface. These were cross-examined to ensure that the exams were identical across the three semesters. Since all questions were identical across semesters, all were included in the analysis.

\subsubsection{Survey Results} \label{sec:survey}
At the end of the CS1-CR semester, students were required to take a survey asking questions regarding their AI usage, preparation for code review sessions, and perception of code review sessions. The full set of questions given are shown in Table~\ref{tab:survey_questions}. 
\edit{
Our IRB protocol required all study data to be anonymized by staff in the University's data office, and their anonymization procedure did not allow us to link survey responses to exam results. Thus we are unable to disaggregate survey results by exam or course grade.  
}

\begin{table}[htbp]
    \caption{Survey questions given at the end of the CS1-CR semester. Questions were designed to measure student use of AI and perceptions of course policies.}
    \label{tab:survey_questions}
    \small
    \centering
\begin{tabular}{|p{0.55\textwidth}|p{0.40\textwidth}|}
    \hline
    \textbf{Question} & \textbf{Answer Options} \\
    \hline
    
    How much time did you spend preparing for Code Reviews outside of the work required for the assignment? & 
    $\circ$ Less than 15 minutes \newline
    $\circ$ 15--30 minutes \newline
    $\circ$ 0.5--2 hours \newline
    $\circ$ Greater than 2 hours \\
    \hline
    
    How did you prepare for the code reviews? (Check all that apply) & 
    $\square$ Rereading my code \newline
    $\square$ Visiting the coaching center \newline
    $\square$ Practicing concepts \newline
    $\square$ Watching the videos \newline
    $\square$ Writing/revising plan \newline
    $\square$ Other: \underline{\hspace{2cm}} \\
    \hline
    
    In what ways did you use AI tools while working on the programming assignments? & 
    $\square$ Write code for me \newline
    $\square$ Debug my code \newline
    $\square$ Explain code to me \newline
    $\square$ Explain concepts to me \newline
    $\square$ Other: \underline{\hspace{2cm}} \\
    \hline
    
    How often did you use AI tools while working on the assignments for this course? & 
    $\circ$ Never \newline
    $\circ$ On some assignments \newline
    $\circ$ Once per assignment \newline
    $\circ$ Multiple times per assignment \\
    \hline
    
    Did the way you prepared for code reviews change over the course of the semester? If so, how? & \textit{Text Response} \\
    \hline
    
    Did the way you used AI tools change over the course of the semester? If so, how? & \textit{Text Response} \\
    \hline
    
    What did you like about the code reviews? & \textit{Text Response} \\
    \hline
    
    What did you not like about the code reviews? & \textit{Text Response} \\
    \hline
    
    \multicolumn{2}{|c|}{\textbf{Likert Scale}} \\
    \multicolumn{2}{|c|}{\small (1=Strongly Disagree, 5=Strongly Agree)} \\
    \hline
    The code reviews motivated me to understand my code better. & 1 \quad 2 \quad 3 \quad 4 \quad 5 \\
    \hline
    The code reviews helped me to avoid over-reliance on AI tools. & 1 \quad 2 \quad 3 \quad 4 \quad 5 \\
    \hline
    This course should continue to use code reviews. & 1 \quad 2 \quad 3 \quad 4 \quad 5 \\
    \hline
    My grades on the code reviews accurately reflect my understanding. & 1 \quad 2 \quad 3 \quad 4 \quad 5 \\
    \hline
    My grades on the exams accurately reflect my understanding. & 1 \quad 2 \quad 3 \quad 4 \quad 5 \\
    \hline
    My grades on the quizzes accurately reflect my understanding. & 1 \quad 2 \quad 3 \quad 4 \quad 5 \\
    \hline
\end{tabular}
\end{table}

\subsubsection{Keystroke Logging Data} \label{sec:keystroke}
To estimate how student AI usage might be impacted by new policies, we determined what percentage of student code was pasted into the IDE during assignments over CS1-CR compared to previous semesters. Fa24 and CS1-CR students were required to use the ShowYourWork~\cite{showyourwork} plugin in their PyCharm IDEs through all of their coding assignments. This plugin logs each keystroke made in the IDE editor, including which characters were typed and a timestamp. Other edit events, such as cut, paste, accepts of code suggestions, and changes from refactoring tools, were also logged. This keystroke data was deidentified by authorized university data handlers, which included removal of the specific characters typed at each event, but still ensured that the research team could see the total number of characters typed, pasted, or otherwise modified at each event. 
Our keystroke analysis is performed over eight assignments from Fa24 and CS1-CR. Table~\ref{tab: assignments used} gives a complete list of all assignments given over Fa23, Fa24, and CS1-CR, and provides an explanation for our limited selection of assignments to compare. \edit{Keystroke logs were not collected in Fall 2023.} Table~\ref{tab:assignment description} provides a brief description of each assignment's requirements. 


\begin{table}[htbp]
\centering
\caption{Descriptions of the programming assignments used.}
\begin{tabular}{c p{0.85\textwidth}}
\toprule
Assignment & Description \\
\midrule
A1a  & Using functions from a simple drawing library, draw a snowman \\

A1b  & Take values given in user input to draw a custom image with a simple drawing library \\

A2a  & Write mathematical formulas that take user input; Write simple functions \\

A2b & Use input validation, data type conversion, and conditional statements to make decisions \\
                
A3a & Use logical operators, lists, and basic object-oriented programming to identify leap year \& creature \\

A3b & Use of object oriented programming, loops to create pyramid and math game with animated score \\

A3b' & Use of object oriented programming, loops to create animated score on thermometer\\

A4 & Use of user-defined functions, if-elif logic, loops to draw customizable chessboard \& geometric patterns \\

A4a & Write a software development plan. Use of nested loops \& string formatting to create a pyramid\\

A5a & Implement a game using pygame library with handling of game loops, keyboard movement, collision detection, and clickable change in character expression \\

A5a' & Implement a game using pygame library tracking object states, keyboard movement, collision detection \\

A5b & Implement a game using pygame library with cursor movement, collision detection, timed rounds \\

A5b' & Implement game using pygame library with cursor movement, collision detection, timed rounds and clickable change in character expressions\\

A6 & Two-player game using pygame library handling game loops, win condition, keyboard movement,  classes, audio, and images \\

A7 & Basic search and sort algorithms, overloaded operators \\
\bottomrule
\end{tabular}
\label{tab:assignment description}
\end{table}

\subsection{Data Analysis}

\subsubsection{Quantitative Comparison of Test Scores} 
To determine how student performance might have been impacted by CS1-CR's new policies, we ran pairwise $t-$tests comparing the Fa23, Fa24, and CS1-CR exam scores. The dataset used in this portion of our analysis is described further in Section~\ref{sec:scores}.

\subsubsection{Keystroke Analysis Methods} \label{sec:keystroke methods}
Our keystroke analysis is done primarily to discover whether student pasting behavior is consistent with the hypothesis that students are using AI more in the CS1-CR course than in Fa24. That is, since CS1-CR allows students to use AI, is there evidence in the keystroke logs that they are, in fact, using \edit{more} AI \edit{than before the policy change}? We use three primary statistics to answer this question: number of pastes per submission, total number of characters pasted per submission, and percentage of characters that were pasted relative to the total number of characters inserted into the file. For simplicity, we use the term ``paste'' to comprise any edit event that inserts more than two characters, including code copied from an external AI tool and acceptance of inline code suggestions from an AI. Pastes of exactly two characters are not included in order to avoid the case when the programmer types an open parenthesis/bracket and the PyCharm IDE automatically inserts a close parenthesis/bracket in the same event.

We computed two measures of student effort on assignments: time-on-task (time taken) and number of keystrokes to complete assignments. Time-on-task is computed by summing elapsed time between pairs of keystrokes. Elapsed times greater than 5 minutes are discarded as the students were probably disengaged~\cite{hart2023accurate}.




\subsubsection{Quantitative Survey Methods}
The survey (Table~\ref{tab:survey_questions}) included some multiple choice, select all that apply, and Likert scale questions. The aggregate results for each of these questions were processed to display the distribution of answers from all students. 

\subsubsection{Qualitative Survey Methods}
For each of the free-response questions in the survey (Table~\ref{tab:survey_questions}), the first and last authors individually determined common themes found throughout the student responses.
\edit{After coding about 30 responses to each question, they met to compare their codebooks, resolve any disagreements, and merge common themes to come up with a combined codebook. This process was repeated several times until both authors had coded the entire dataset and reached agreement on every code application. Each time a new code was added, both coders would go back over the entire data set to apply new codes to anywhere they had been missed.}
The researchers involved in this process had no connections to students in the CS1-CR course, and no instructors or TAs for the course \edit{were involved in reviewing the data.} \edit{The final codes were} then processed to display the frequency of each code to measure student perceptions and responsive actions taken in regards to the new policies implemented in CS1-CR.

\section{Results}

\subsection{Quantitative Comparison of Test Scores}
\begin{figure}[htbp]
    \centering
    \includegraphics[width=\linewidth]{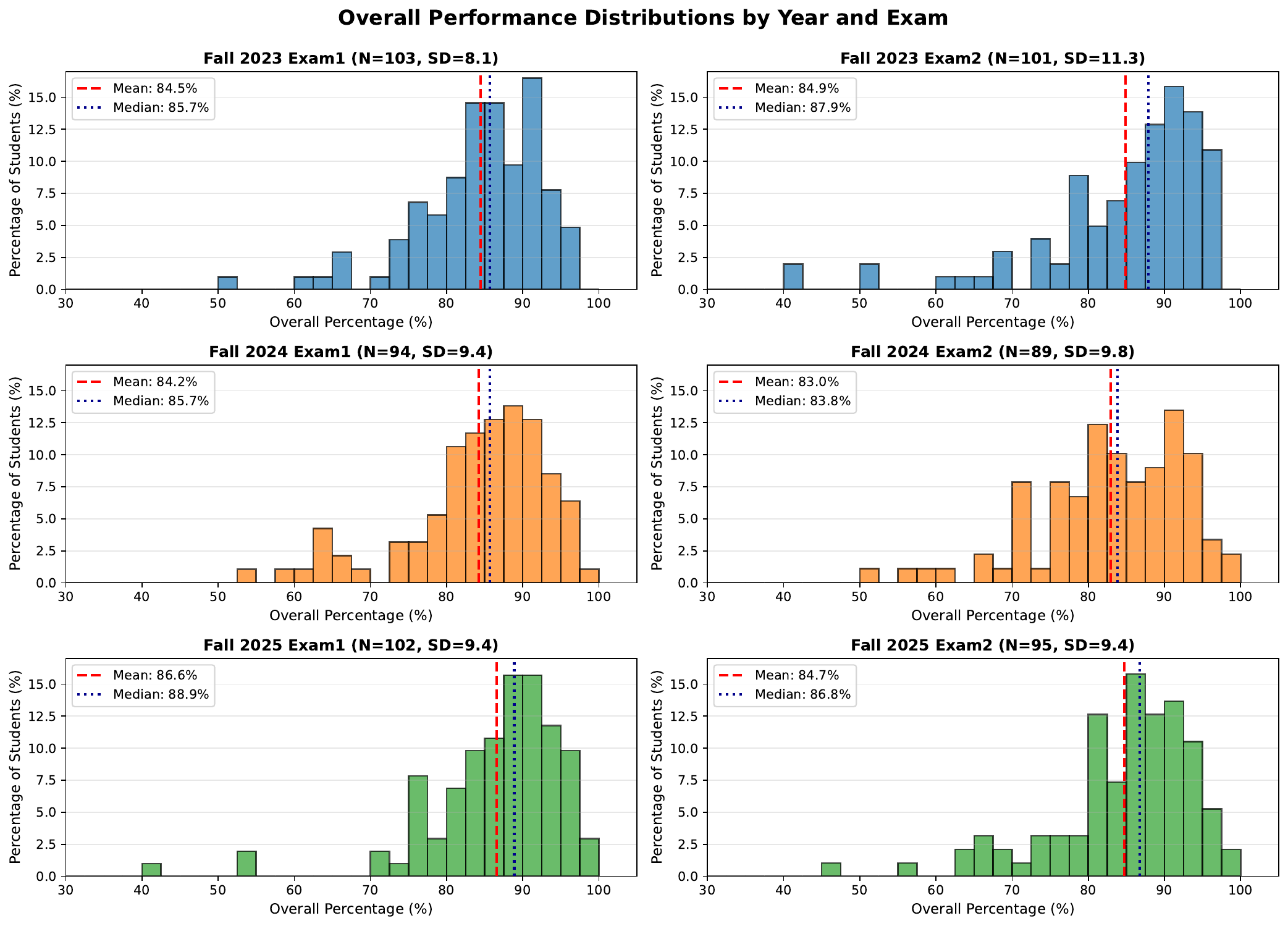}
    \caption{Distribution of exam scores across the Fall 2023, Fall 2024, and CS1-CR semesters. Results on both exams were relatively stable between Fall 2023 and 2024. On Exam 1, CS1-CR students (Fall 2025) performed about 2\% better than the prior semesters. None of the the differences were statistically significant at $\alpha = 0.05$, though some were marginally significant. Results of $t-$tests between these distributions are given in Table~\ref{tab:ttest}.}
    \label{fig:performance}
\end{figure}

The pairwise comparisons for Exam 1 across each semester showed a slight improvement in performance from both the Fa23 and Fa24 courses when compared to CS1-CR. Comparing Exam 2 from Fa23 to CS1-CR showed a 0.15\% decrease in performance, but there was once again a nearly 2\% increase from Fa24 to CS1-CR. None of these performance changes showed any statistical significance. \edit{These results fail to indicate any statistically significant learning gains or losses from the new policies adopted in CS1-CR.} Table~\ref{tab:ttest} showcases these results. Figure~\ref{fig:performance} showcases the score distributions for each exam across the semesters.

\begin{table}[htbp]
    \caption{Pairwise $t-$test results across exams taken over the Fa23, Fa24, and CS1-CR semesters. Results on both exams were relatively stable between Fa23 and Fa24. On Exam 1, CS1-CR students performed about 2\% better on average than the prior semesters. None of the differences were statistically significant if using a cutoff of $\alpha = 0.05$. These distributions and their means are visualized in Figure~\ref{fig:performance}.}
    \label{tab:ttest}
    \centering
    \begin{tabular}{c||c|c|c|c}
        
        & Comparison &  \% Score Increase  & $t$ & $p$  \\\hline \hline
        Exam 1 & Fa23 vs. Fa24 & -0.30\% & -0.242 & 0.809  \\\hline
        Exam 1 & Fa23 vs. CS1-CR & 2.08\% &  1.697 & 0.091 \\\hline
        Exam 1 & Fa24 vs. CS1-CR & 2.38\% & 1.772 & 0.078  \\\hline \hline
        Exam 2 & Fa23 vs. Fa24 & -1.92\% & -1.245 & 0.214 \\\hline
        Exam 2 & Fa23 vs. CS1-CR & -0.15\% & -0.104 & 0.917  \\\hline
        Exam 2 & Fa24 vs. CS1-CR & 1.76\% & 1.248 & 0.213 \\
    \end{tabular}
\end{table}

\subsection{Keystroke Analysis Results}

\begin{figure}[]
    \centering
    \subfloat[]{
      \includegraphics[width=.48\linewidth]{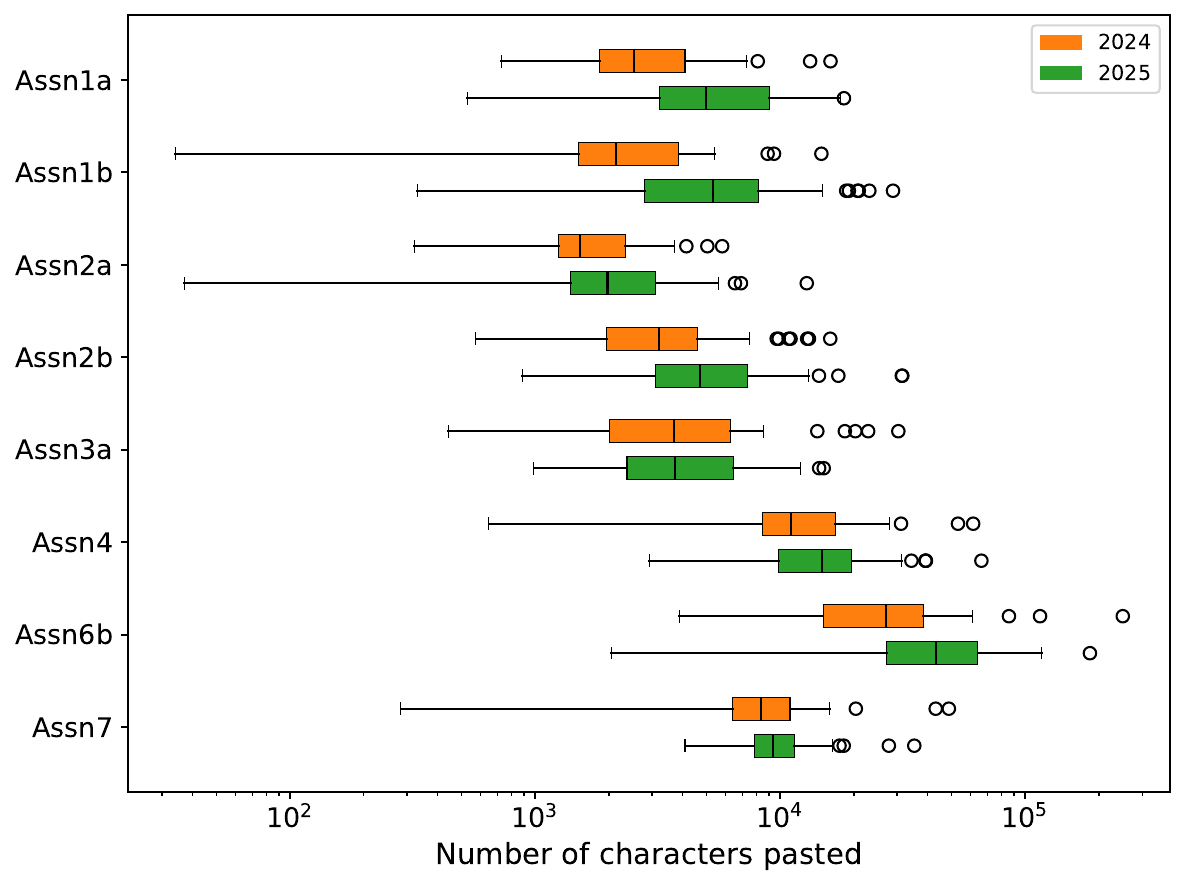}
      \label{subfig:Paste size per submission}
    }\hfill
    \subfloat[]{
      \includegraphics[width=.48\linewidth]{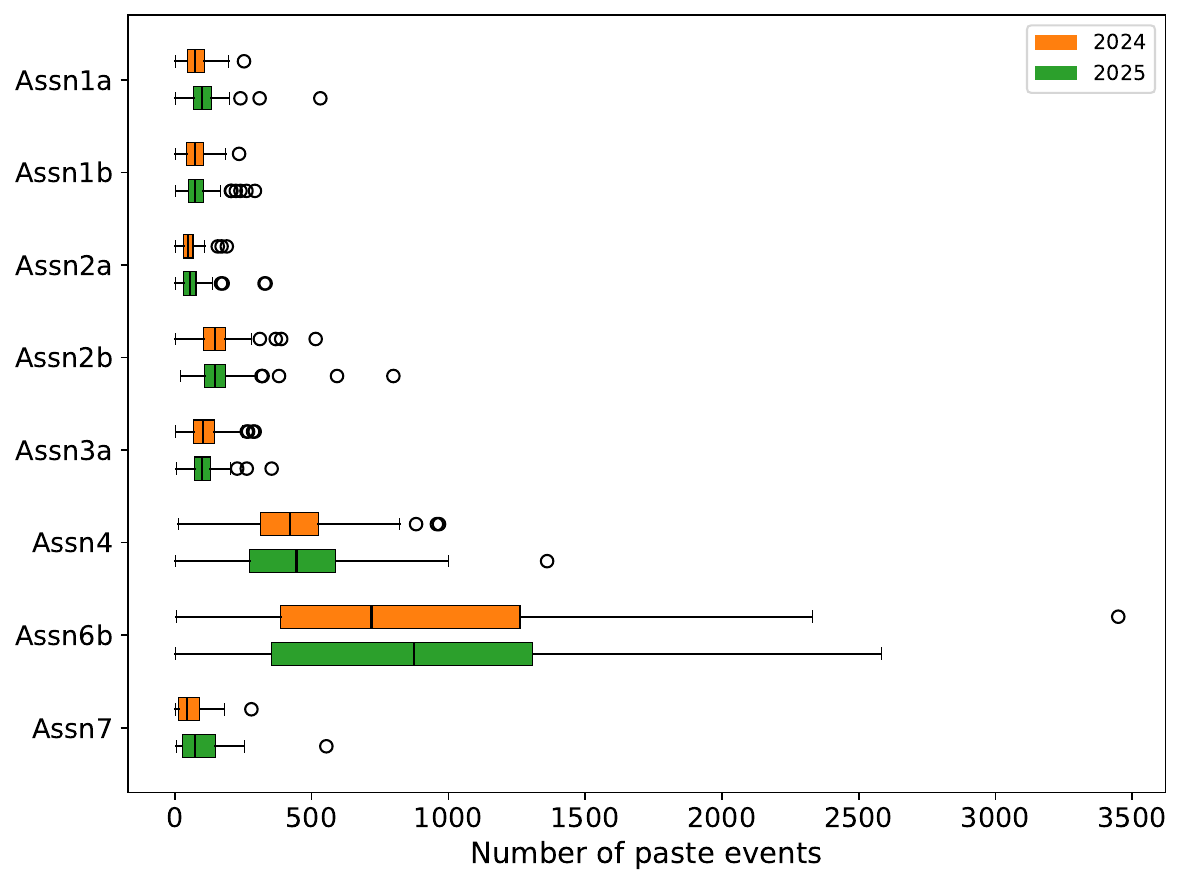}
      \label{subfig:Paste events}
    }
    
    \caption{Difference in paste events in coding assignments between the Fall 2024 and CS1-CR semesters. \protect\subref{subfig:Paste size per submission} Distribution of the total number of characters pasted in each submission. Assignments A1a, A1b, and A6b show large increases in total number of characters pasted per submission. \protect\subref{subfig:Paste events} Distribution of number of paste events per submission across Fall 2024 and CS1-CR assignments. While the difference in number of pastes varies by assignment, the aggregate number of paste events per submission across assignments appears to remain relatively consistent between years.}
    \label{fig:keystroke data}
\end{figure}

\begin{figure}[]
    \subfloat[]{
      \includegraphics[width=.3\linewidth]{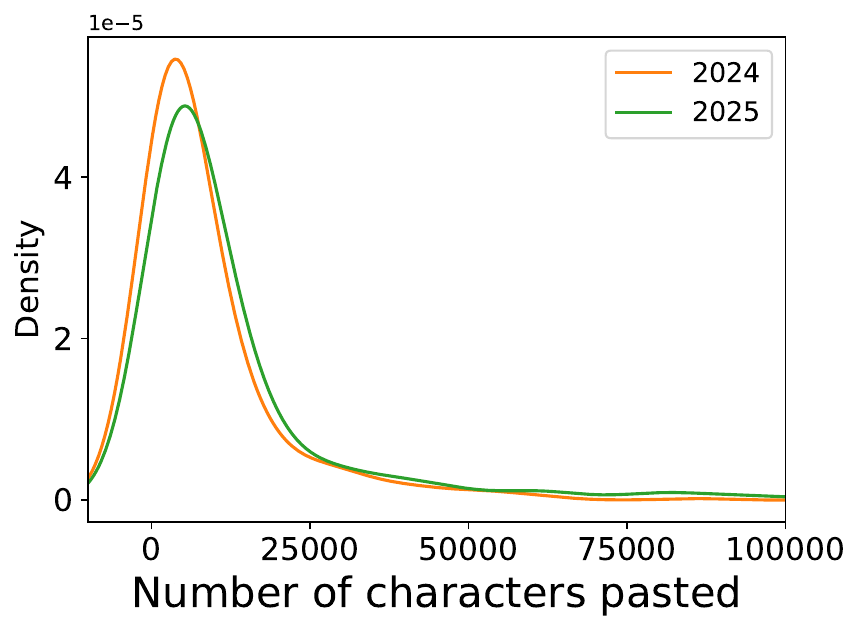}
      \label{subfig:kde_paste_size_per_submission}
    }
    \subfloat[]{
      \includegraphics[width=.3\linewidth]{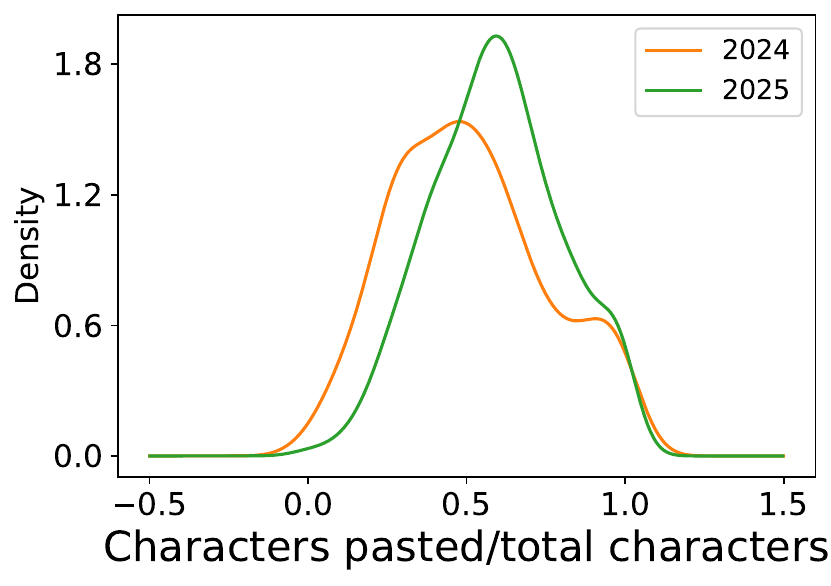}
    \label{subfig:kde_characters_pasted_per_total_characters}
    }
    \\
    \subfloat[]{
      \includegraphics[width=.3\linewidth]{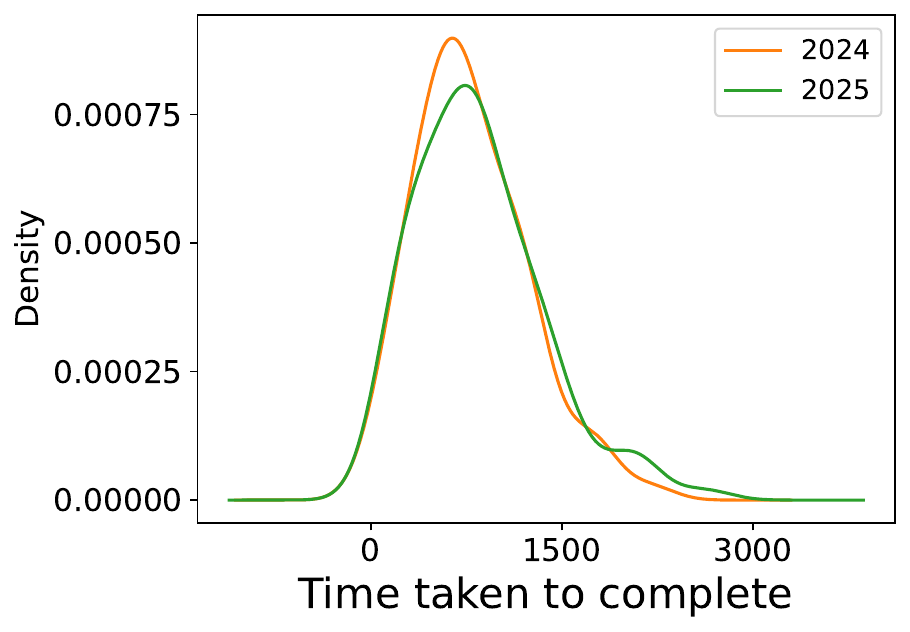}
      \label{subfig:kde_completion_time}
    } 
    \subfloat[]{
      \includegraphics[width=.3\linewidth]{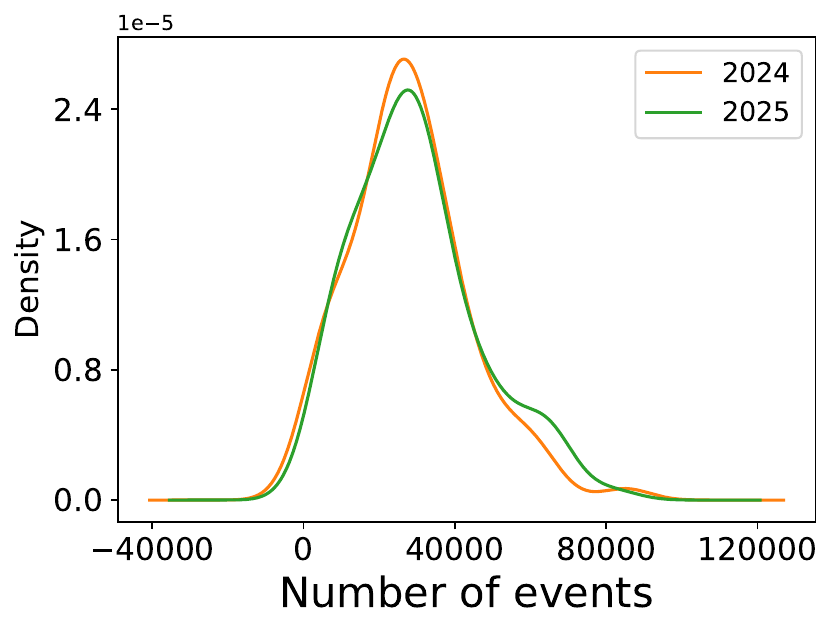}
      \label{subfig:kde_events_per_student}
    } 
    \caption{\protect\subref{subfig:kde_paste_size_per_submission} Distribution of number characters pasted in each submission. \protect\subref{subfig:kde_characters_pasted_per_total_characters} Distribution of percentage of characters pasted per total characters in each submission. \protect\subref{subfig:kde_completion_time} Distribution of the time taken per student across all assignments. \protect\subref{subfig:kde_events_per_student} Distribution of number of events per student.
    Subfigures (a) and (b) show that students in CS1-CR had more paste events. Subfigures (c) and (d) show that students took the same amount of time and keystrokes to finish their assignments across semesters.
    }
    \label{fig:kde_plots_paste_behaviors}
\end{figure}

Figure \ref{fig:keystroke data} shows the distribution of number of characters pasted and number of paste events for Fa24 and CS1-CR. We see in Figure~\ref{subfig:Paste size per submission} that the number of characters pasted per submission generally went up between Fa24 and CS1-CR. For each submission across all assignments, the number of characters pasted went up $(U=87770, p<0.0001)$ 
(Figure~\ref{subfig:kde_paste_size_per_submission}). It could be argued that this increase is simply due to an increase of total number of characters input during the programming process for some reason other than \edit{increased} AI usage. However, the percentage of characters pasted relative to total number of characters input $\left(\frac{\#pasted}{\#pasted+\#typed}\right)$ increased from $61.0\%$ in Fa24 to $68.1\%$ in CS1-CR. A Mann-Whitney U test on the two distributions of submission-level percentages shows that this increase is significant $(U=85517, p<0.0001)$ (Figure~\ref{subfig:kde_characters_pasted_per_total_characters}). 
A two proportion $z-$test might appear more appropriate for the proportion of pasted characters relative to total input. However, characters within a submission are not independent as a single paste event can contribute hundreds of characters, violating the independence assumption of the two proportion $z-$test. 
This means that paste behavior increased independently of other programming process behaviors, likely due to increased AI usage.

Figure~\ref{subfig:Paste events} shows that the number of paste events appears to have changed little from Fa24 to CS1-CR. Aggregated across usages, there is only very weak evidence of change ($U=102388, p=0.10)$. The increase in characters pasted with unchanging number of pastes indicates that students appear to be pasting larger blocks of AI-generated code. While not a finding central to our work in this paper, it does gives some idea of how students are using the AI systems, that is, prompting for and pasting larger pieces of code rather than interacting with smaller sections.

Figures~\ref{subfig:kde_completion_time} and~\ref{subfig:kde_events_per_student} show distributions of time-on-task and number of keystrokes for each submission. We expected that, with the increase in pastes, presumably from increased AI use, that student effort on assignments would go down, resulting in a decrease in both number of keystrokes and time-on-task. Surprisingly, we found no evidence of this with $(U=4336, p=0.53)$ for time-on-task and $(U= 4372, p=0.60)$ for number of keystrokes (Figures~\ref{subfig:kde_completion_time} and~\ref{subfig:kde_events_per_student}, respectively). It appears that students engaged with the code just as much when AI use was allowed as when it was not. We hypothesize that this was because students were motivated to prepare for code reviews.

\subsection{Quantitative Survey Results}
\begin{figure}
    \centering
    \includegraphics[width=\linewidth]{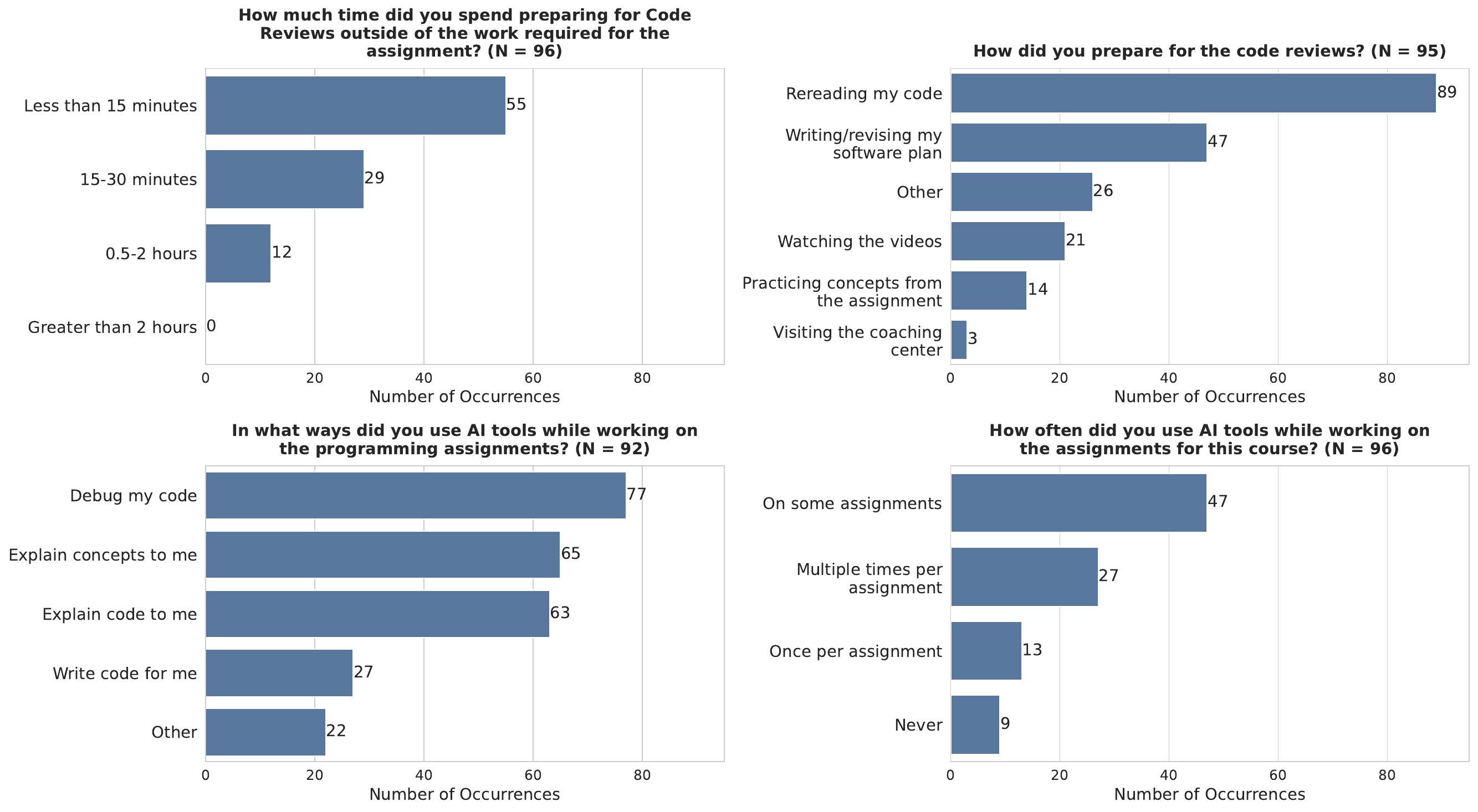}
    \caption{Results from multiple choice survey questions distributed to CS1-CR students at the end of the CS1-CR semester. }
    \label{fig:multiple choice}
\end{figure}

Responses to the multiple-choice questions (Figure~\ref{fig:multiple choice}) show that students mostly prepared for code reviews by rereading their code and writing/revising their software plans. While working on assignments, AI tools were mostly used for debugging code and explaining concepts/code. AI tools were used in these ways on some assignments, with 27 students indicating that they were used multiple times per assignment. 

\begin{figure}
    \centering
    \includegraphics[width=\linewidth]{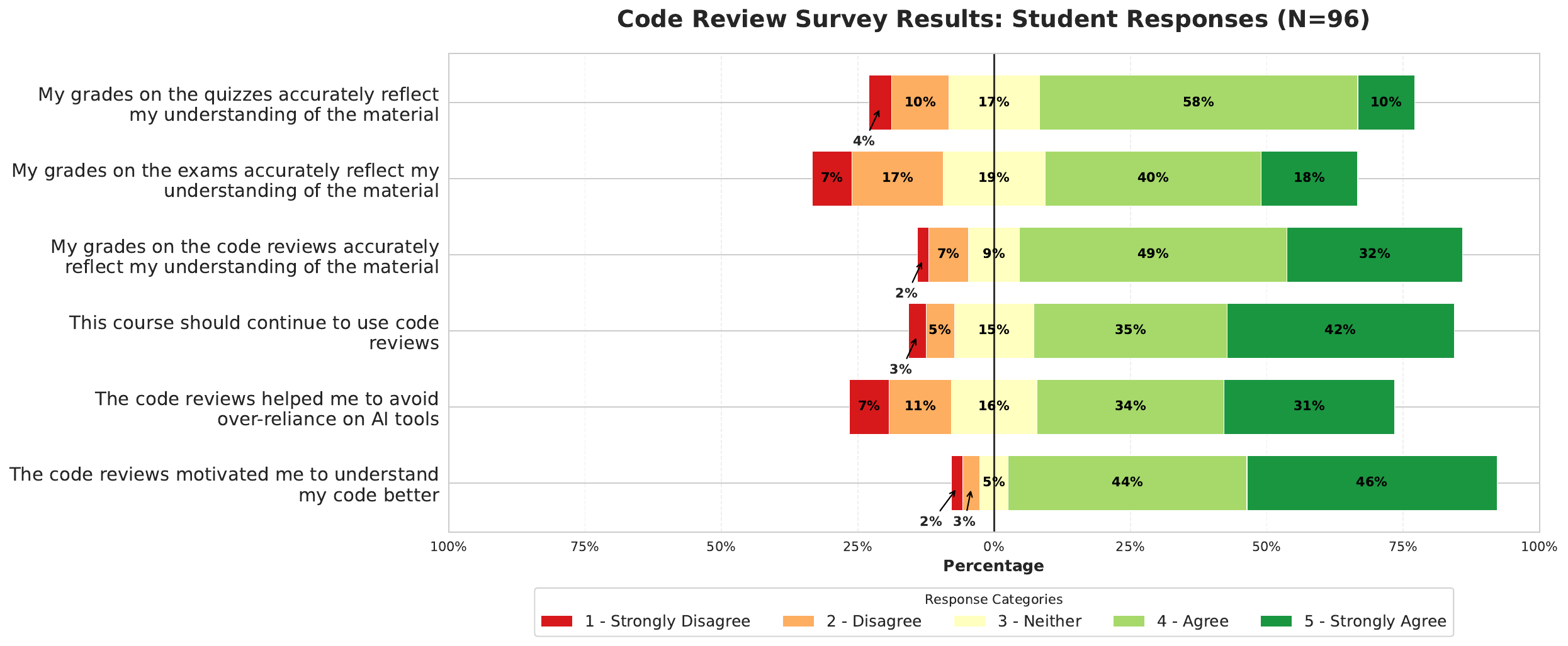}
    \caption{Results from Likert scale survey questions distributed to CS1-CR students at the end of the CS1-CR semester. These results show overwhelmingly positive sentiment towards CS1-CR, strongly indicating that they helped facilitate deeper understanding and avoidance of over-relying on AI tools.}
    \label{fig:likert}
\end{figure}

The Likert scale questions (Figure~\ref{fig:likert}) show overwhelmingly positive sentiment towards the code reviews, with students indicating that their grades on the code reviews accurately reflect their understanding of the material more than their grades on the quizzes or exams. Additionally, 65\% of students agreed that code reviews helped to avoid over-reliance on AI tools, and 90\% reported being motivated to understand their code better. Most students also agreed that the course should continue to use code reviews.

\subsection{Qualitative Survey Results}

Our qualitative analysis of students' written feedback reflects the positive sentiment shown in the quantitative results. The following sections discuss themes identified for each free-response question. 

\subsubsection{Did the way you prepared for code reviews change over the course of the semester? If so, how?} \label{sec:prep}

\begin{table}[t]
\small
\setlength{\tabcolsep}{3pt}
\caption{Did the way you prepared for code reviews change over the course of the semester? ($N=91$)}
\label{tab:prep_changes}
\centering
\begin{tabular}{p{0.25\linewidth} p{0.60\linewidth} l}
\toprule
\textbf{Theme} & \textbf{Description} & \textbf{Count (\%)} \\
\midrule
More Time & The student generally spent more time preparing for code reviews over time & 45 (49.45\%) \\
No & The student's way of preparing for code reviews had no changes over time & 35 (38.46\%) \\
More Reading Code & The student spent more time reading and attempting to understand their code & 23 (25.27\%) \\
Less Time & The student generally spent less time preparing for code reviews over time& 9 (9.89\%) \\
More planning/working on code plan & The student spent more time intentionally planning out their code during the process of completing the assignment & 8 (8.79\%) \\
Used AI More & The student used AI tools to prepare & 5 (5.49\%) \\
Practicing Self-Explanation & The student spent time practicing how they would explain concepts and implementation details in their code to a TA & 3 (3.30\%) \\
Identifying possible questions & The student put more effort into thinking about the possible questions that a TA might ask in their next review session and preparing to respond to them & 3 (3.30\%) \\
\bottomrule
\end{tabular}
\end{table}
Table~\ref{tab:prep_changes} gives a detailed description and the number of occurrences found for each theme identified for this question.
Of the 91 respondents to this question, 45 students indicated that they generally spent more time preparing for the code review sessions as the semester progressed, and 35 indicated that they had no change in how they prepared. Interestingly, 9 students said they did less preparation as the semester progressed. There are several notable reasons reported that explain this decrease in preparation effort, including:

\begin{itemize}[topsep=0pt, partopsep=0pt, itemsep=0pt, parsep=0pt]
    \item Improved confidence and comfort with the code reviews: ``\textit{If anything, I spent less time preparing for the code reviews as the semester went on because I started feeling more comfortable with them.}''
    \item Becoming more involved in trying to understand the concepts while working on the assignment: ``\textit{I prepared a little less in a way because I noticed that as I was writing my code I made sure I understood it better so while preparing for the code review I didn't need to worry as much because I knew I understood my code.}''
    \item Forgetting to prepare: ``\textit{Towards the end I forgot to go over my code beforehand which made it a bit harder to answer the questions.}''
\end{itemize}

\subsubsection{Did the way you used AI tools change over the course of the semester? If so, how?} \label{sec:ai_changes}

\begin{table}[t]
\small
\setlength{\tabcolsep}{3pt}
\caption{Did the way you used AI tools change over the course of the semester? If so, how? ($N=89$)}
\label{tab:ai_changes}
\centering
\begin{tabular}{p{0.25\linewidth} p{0.60\linewidth} l}
\toprule
\textbf{Theme} & \textbf{Description} & \textbf{Count (\%)} \\
\midrule
More for Debugging & More AI usage for assistance with identifying and fixing bugs  & 31 (34.83\%) \\
More for Understanding & More AI usage for help with explaining challenging concepts & 30 (33.71\%) \\
No & No change in how AI was used through the semester & 23 (25.84\%) \\
More as concepts got harder & More AI usage as concepts became more challenging & 18 (20.22\%) \\
More for code quality/revision & More AI usage to improve already functioning code & 6 (6.74\%) \\
More Generating Code & More AI usage for code generation & 5 (5.62\%) \\
Less Generating Code & Less AI usage for code generation & 3 (3.37\%) \\
Less for Understanding & Less AI usage for help with understanding concepts & 3 (3.37\%) \\
More for Planning & More AI usage to help plan out their code when starting a new assignment & 2 (2.25\%) \\
More Hinting & More AI usage to provide hints towards the next steps in implementation & 1 (1.12\%) \\
\bottomrule
\end{tabular}
\end{table}
Table~\ref{tab:ai_changes} gives a detailed description and the number of occurrences found for each identified theme. 
Of the 89 respondents to this question, 31 indicated a shift towards using AI tools more for debugging assistance as the semester progressed: ``\textit{I first used it to explain the code I wrote to help for the first code reviews, but when the assignment got more difficult, I used it to debug primarily}''. 18 students mentioned that they used AI to explain challenging concepts as they became increasingly complex: ``\textit{didn't use AI at first, but as subjects got more complicated, I used AI to better understand what was going on}''; However, three students indicated a reduction in this usage by shifting to other use cases instead: ``\textit{AI became more of a debugging tool rather than a teacher for concepts.}'', becoming aware of AI biases and hallucinations: ``\textit{I realized that AI was frequently getting me down rabbit holes I did not need to go down.}'', and accessing high-quality course materials: ``\textit{At first I used it to explain certain quirks about Python, but I realized that re-watching a section of the videos in Canvas gave me more than enough explanation if I reviewed them a handful of times.}''.

\subsubsection{What did you like about the code reviews?} \label{sec:likes}
\begin{table}[t]
\small
\setlength{\tabcolsep}{3pt}
\caption{What did you like about the code reviews? ($N=94$)}
\label{tab:likes}
\centering
\begin{tabular}{p{0.25\linewidth} p{0.60\linewidth} l}
\toprule
\textbf{Theme} & \textbf{Description} & \textbf{Count (\%)} \\
\midrule
Learning & Code reviews provided opportunities and/or motivation to pursue deeper learning and understanding & 58 (61.70\%) \\
Interacting With TAs & Positive experiences interacting with TAs. & 22 (23.40\%) \\
Fair Grading & Code reviews provided a more fair way to grade coding assignments; they appreciated having opportunities to justify their coding decisions and express their understanding rather than grading occurring behind closed doors. & 16 (17.02\%) \\

Feedback & Receiving personalized feedback on code and understanding of the assignment learning objectives & 12 (12.77\%) \\
Reduced AI Reliance & Reduced motivation to rely on AI tools for coding assignments & 9 (9.57\%) \\
Scheduling & The scheduling process generally went smoothly & 5 (5.32\%) \\
\bottomrule
\end{tabular}
\end{table}
Table~\ref{tab:likes} gives a detailed description and the number of occurrences found for each theme identified for this question.
Of the 94 respondents, 58 students discussed CS1-CR enhancing their learning experience. Knowing that they would have to explain their solutions and understanding of concepts in the future provided a strong incentive to engage in deeper learning while working on assignments:
\begin{quote}
    ``\textit{It was an opportunity to be prompted to understand my code better, rather than just turning in code that worked, I needed to more fully understand why it worked, and being able to explain it deepens my understanding}'' 
\end{quote}
Several students also learned from new perspectives offered by TAs conducting code reviews: 
\begin{quote}
``\textit{I liked that the TAs were able to explain how things work and why they work to me. If I didn't finish an assignment, they would give me a brief explanation of how I might have improved on something or how to better implement things into my code.}''
\end{quote}

Nine students also reported that the added incentive to understand their code reduced their reliance on AI tools: ``\textit{With the possibility to use AI to write a somewhat successful program, the code reviews motivated me to use AI only as a help/tool.}''

\subsubsection{What did you not like about the code reviews?} \label{sec:dislikes}

\begin{table}[t]
\small
\setlength{\tabcolsep}{3pt}
\caption{What did you not like about the code reviews? ($N=92$)}
\label{tab:dislikes}
\centering
\begin{tabular}{p{0.25\linewidth} p{0.60\linewidth} l}
\toprule
\textbf{Theme} & \textbf{Description} & \textbf{Count (\%)} \\
\midrule
TA inconsistency (difficulty) & Some TAs would grade more harshly than others or ask more difficult questions & 31 (33.70\%) \\
Scheduling difficulties & Scheduling a code review session was difficult and inconsistent & 20 (21.74\%) \\
Too Much Weight on Grade & Grade weight on code reviews had too high of an impact on overall assignment scores; Getting no points for failing to schedule a session was too high a cost. & 14 (15.22\%) \\
TA inconsistency (time) & Some TAs would extend the duration of the code review sessions far beyond the 15-minute window & 13 (14.13\%) \\
Inconvenient & It was not convenient to attend the code review sessions every week & 6 (6.52\%) \\
Seemed Unnecessary & The student believed the code reviews were not a necessary part of their learning & 5 (5.43\%) \\
Rude TA & Negative experience(s) with TA(s) being rude, belittling, or disrespectful & 3 (3.26\%) \\
\bottomrule
\end{tabular}
\end{table}
Table~\ref{tab:dislikes} gives a detailed description and the number of occurrences found for each theme identified for this question.
The majority of dislikes reported by students align with those discussed in prior work on oral exams---\citet{gharibyan2005oral} discusses the possibility for the one conducting the exam to do so unfairly, or for students to believe they were treated unfairly. In our survey, 31 of the 92 respondents reported instances of some TAs grading more harshly than others or asking questions beyond the scope of the assignment. Despite each code review session having a 15-minute time limit, 13 students reported issues with this time constraint not being respected by graders, with some even going over an hour long. Three respondents also reported interactions with a TA who made them feel disrespected or belittled. The following student quote summarizes all of these issues:
\begin{quote}
    ``\textit{Most of the TAs were awesome and would give me the advice I needed and the grade I thought I deserved. There was 1 TA who would go off script in the code reviews, he would ask his own questions that didn't make a lot of sense, he made me feel dumb on multiple occasions instead of showing me what I did wrong then giving me advice he would just point it out and look at me like I was dumb. And since he made up his own questions the code reviews went extra-long. I had to miss class one day because I had a code review that lasted around an hour and 20 minutes. I feel like all the TAs should have to conduct a code review in the exact same way just to be fair.}''
\end{quote}

In Section~\ref{sec:context}, we discuss that students were required to schedule these sessions through the university's tutoring center within 48 hours of the submission deadline for each assignment. 20 students reported issues with these scheduling constraints, wishing the window was larger than two days, disliking their weekly frequency, or having no open time slots that aligned with their needs. 

Section~\ref{sec:context} also discusses how homework assignments were graded with 70\% of the score going towards code reviews and 30\% going to the code submission alone, or an indisputable zero if they failed to schedule one for an assignment. 14 students reported that they disliked these grading policies. Many from this group specifically mentioned disliking getting no points if they forgot to schedule a code review, even if they had spent hours coding their submission: 

\begin{quote}
    ``\textit{I did not like how if you missed a code review, you got 0 credit for the assignment... if we spent a lot of time doing the assignment and just forgot to do the code review, yeah, we can miss the points for the code review itself, which is most of them, but the student should still receive the points for the actual assignment part since they spent hours on it.}''
\end{quote}
    
Some did not like how the rubrics were designed, feeling like they sometimes deducted too many points for certain implementation details: 

\begin{quote}
``\textit{I had a week where my code was complete, my understanding was there and I could fully explain my logic, code, thought processes, but I put minimal effort into my design plan. That led to an overall deduction in a category that was largely based on understanding the code, which I felt I did entirely.}''
\end{quote}

\section{Discussion}
Because this study used data collected and analyzed in many different ways, giving full answers to the research questions requires synthesizing across the  different methodologies and results. Here, we attempt to distill the main answers to each of the research questions, drawing on conclusions found across the various types of data analysis performed.


\subsection{RQ1: How is student performance impacted by policies of CS1-CR?}
As shown in the survey results the most common sentiment of students, by far, is that the code reviews helped them to learn---a view expressed by over half of students (58/94). 
\edit{These results are consistent with findings in prior work, which also report students claiming to have learned from oral assessments~\cite{dick2005interviews}, or understand the content better because of their oral assessments~\cite{grunwald2015personalized}. In addition, we provide findings on how the students felt the code reviews mitigated effects of AI reliance.}
65\% of students agreed that the code reviews helped them avoid over-reliance on AI tools, with 16\% being neutral and 18\% disagreeing. Some students additionally noted in the open-response questions that the code reviews helped mitigate reliance on AI tools (9/94). 

As shown by the student test scores in Figure~\ref{fig:performance} and Table~\ref{tab:ttest}, student scores \edit{showed no statistically significant improvements or losses under the policise of CS1-CR.}
\edit{Prior studies on oral assessments in CS education do not provide measures of student performance, instead relying on either instructor experiences or data from student survey responses~\cite{grunwald2015personalized, dick2005interviews, ohmann2019oral, ohmann2025oral}.}

\subsection{RQ2: How is student usage of generative AI tools impacted by policies of CS1-CR?}
As shown by the survey question in Figure~\ref{fig:multiple choice}, almost all students used AI tools for at least a part of each assignment. The majority used AI to debug their code and have code explained to them. \edit{Only} 27 out of 92 students surveyed said that they had AI tools write code for them.

The keystroke analysis given in Figure~\ref{fig:keystroke data} gives additional information. Students appear to have increased their usage of generative AI tools when they were explicitly allowed to use them in CS1-CR. Combined with the performance data above, this shows that students were using AI tools more, but were still able to learn just as much as before. 

\edit{While the keystroke logging software does accurately report all paste events, it does not track the source of any pasted text that was not copied within the IDE's text editor. This ensures student privacy when acting outside of their IDE, but adds some ambiguity to our analysis, as we cannot reasonably infer that all pasted text was copied from AI output. Although previous semesters explicitly forbade AI usage for coding assignments, these policies are notoriously difficult to enforce~\cite{bernstein2025beyond}. However, prior work indicates a decline in usage of community-driven question forums, alongside an increase in GenAI preference for code generation and assistance~\cite{umair2026decline}. With this as evidence, we can reasonably infer that a decent proportion of pasted code from past semesters is AI-generated, and that the increase in pasting we saw in the keystrokes represents an increase in AI usage, not a change from no use to some use.}

\subsection{RQ3: How do students study and prepare for code review interviews in CS1-CR?}
As shown in Figure~\ref{fig:multiple choice}, most students spent less than 15 minutes preparing for code reviews. Our qualitative analysis of free-response questions shows they prepared mostly by rereading their code, with some revisiting their software plan or by otherwise reviewing concepts and ideas from the assignments (Section~\ref{sec:prep} and Table~\ref{tab:prep_changes}).

\edit{While some prior work discusses similar implementation of formative oral assessments~\cite{grunwald2015personalized, dick2005interviews}, we could not find any existing work that provided insight into what efforts students would take to prepare for such assessments, making this a unique contribution of our work.}

\subsection{RQ4: What are the students' perceptions of \edit{code review interviews} in CS1-CR?}
As a whole, students' views of the code reviews were overwhelmingly positive, with 
72\% agreeing, 15\% neutral, and only 8\% disagreeing that the course should continue to use code reviews (Figure~\ref{fig:likert}). Students broadly felt that the code reviews helped them to learn, and that interacting with the teaching assistants was a beneficial experience (Section~\ref{sec:likes} and Table~\ref{tab:likes}). \edit{This is consistent with student feedback from a course that employed similar formative oral exams~\cite{grunwald2015personalized}.}
They also felt that the code reviews were fairly graded, with 81\% of students agreeing that their grades accurately represented their understanding---a higher rate of agreement than for quizzes or exams.

Students' most common grievances with CS1-CR were challenges involving TA inconsistency with grading, time taken on code reviews, and general misconduct towards students. Other common issues were how scheduling and grade weights were handled (Section~\ref{sec:dislikes} and Table~\ref{tab:dislikes}). \edit{Prior work displays similar challenges with varying levels of TA harshness and warmth towards students, with some showing more harsh scrutiny than others, and being condescending towards students. Prior work also indicates that students faced challenges with scheduling conflicts in courses employing similar oral exams.~\cite{grunwald2015personalized, dick2005interviews}} We feel that each of these issues is not a fundamental flaw with CS1-CR's philosophy---rather, they indicate a need for future improvements in TA training and adjustments to scheduling and grading policies. As of writing this paper, CS1-CR has already considered many of these issues and made adjustments accordingly. 
\edit{Specifically, in response to student concerns, the course has: 
(1) eliminated the zero-grade policy, replacing it by a policy where students may still receive up to 30\% of the assignment grade if they miss the interview, (2) made a more thorough effort to achieve consistency in both time and difficulty of code reviews between TAs, and (3) made scheduling more flexible to allow student to complete their code review starting a few days before the assignment is due and extending to a few days afterward.}
Future research is needed to evaluate these adjustments.  

Though we didn't specifically ask about if students agreed with the course policies allowing them to use AI tools on their assignment, students expressed their appreciation for being able to use AI tools in both the quantitative and qualitative portions of the survey. In particular, a high percentage of students used AI tools for debugging and to help their understanding, and did so increasingly over the course of the semester (Table~\ref{tab:ai_changes} and Figure~\ref{fig:multiple choice}).

\section{Limitations}
\label{sec:limitations}

\subsection{\edit{Confounding Factors}}
\label{sec:confounding_factors}
Because code reviews add extra contact time between students and teaching assistants, it is not clear how to construct a true control condition \edit{with no confounding factors} for a study on the effectiveness of code review interviews. \edit{The absences of a true control condition} necessitates the use of quasi-experimental rather than truly controlled studies. In our version of CS1-CR, we gave up one class day each week for code reviews, necessitating some change in instructional practice. While our study design switched to a flipped classroom, future work could explore other alternatives such as skipping some topics during class time or covering lecture material at an accelerated pace on the remaining days. Another alternative would be to keep the same amount of class time while adding interviews, but this would increase the overall contact time between students and course staff, thus \edit{confounding code reviews with increased instructor contact time, rather than with the flipped classroom}. \edit{In any case, our field study was impossible to conduct without introducing an additional variable, whether it's pedagogical differences, increased contact time, or reduction of topics.}
\edit{Despite confounding factors,} CS1-CR provides a proof of concept of a course structure where students are able to use AI tools for their assignments, and still demonstrate similar learning gains to prior semesters. \edit{Future work should examine if these results also replicate with alternative instructional strategies.} We also demonstrated strong positive student sentiment toward the policies of CS1-CR, with almost all students agreeing the course should continue to run under these policies.

\edit{As with all educational studies, replication is needed, but we theorize that our results for RQ1 on student learning will generalize to other courses that choose to implement flipped classrooms with code reviews.  We predict that our results for RQ2, RQ3, and RQ4 on student behaviors and perceptions will generalize to any classroom that implements code reviews, regardless of whether they used a flipped classroom or some other instructional structure. It is less clear if the learning results from RQ1 will replicated under other instructional structures.}

\subsection{\edit{Limited Knowledge of AI Usage}}
Our keystroke analysis work was able to show that students pasted significantly more code when allowed to use AI tools and that they still spent a similar amount of time working on their code in spite of this. However, it does not give insight into the correlation between certain \emph{types} of AI usage and performance on certain types of questions. We leave this more fine-grained analysis to future work. 

\subsection{\edit{Scalability and Ethics}}
While CS1-CR demonstrates a useful course structure for mitigating the harms of AI tools in introductory programming, we do not currently know how this structure may scale to include even larger introductory CS courses or go beyond to incorporate higher-level classes. As of writing this paper, CS1-CR has expanded to cover multiple CS1 courses at the same university, scaling the student population from just under 100 to over 200 students, with the same amount of TAs (11). \edit{As discussed in Section~\ref{sec:related_work_tas}, prior work does not cover TA perceptions of administering oral exams, and so we cannot be sure how the TAs conducting these code reviews might feel about doing them. Future work is needed to examine effects of this scaling on both students and TAs}.

\edit{We attempted to design this survey in a way that would comprehensively identify student preparation, AI usage, and perceptions of code reviews. Our results unveiled some discrepancies in grading fairness that we believe could be solved with better TA training. We acknowledge further ethical concerns that need to be addressed in future work, such as how oral code reviews impact student test anxiety and any potential accessibility concerns for students with disabilities.}

\subsection{\edit{Cohort Concerns}}
\edit{There are valid concerns to be considered when reporting statistics between only a few cohorts. This is due to a phenomenon where one semester's cohort could have been stronger on average than another. This concern is especially prevalent when comparing only two cohorts. This study compares three, with the two cohorts before CS1-CR showing consistent results between one another, and when comparing each of them with CS1-CR. This indicates that this potential concern is likely to not be threatening the validity of our study.}


\section{\edit{Future Work}}
\label{sec:future_work}

Future work will seek to:
    (1) Address student concerns by adjusting policies, grade weights, TA training practices, and scheduling for CS1-CR, 
    \edit{(2) Address potential ethics and accessibility concerns for oral code reviews,}
    (\edit{3}) Identify effects and challenges of scaling CS1-CR policies to larger student populations,
    (\edit{4}) Evaluate TA workloads in and perceptions towards CS1-CR, and
    (\edit{5}) Experiment with pairing code reviews with other instructional models aside from flipped classrooms.

\section{Conclusions}
This paper presented CS1-CR, a course with new policies requiring oral code review sessions conducted by teaching assistants, with a flipped classroom model used to make up for the resulting lost instructional time.
We evaluated student performance, AI usage, and student perceptions of policies in CS1-CR. The results of our analysis demonstrate how these code reviews are capable of combating the harms to cognitive development found through generative AI over-reliance. In spite of several criticisms and issues, students displayed overwhelmingly positive sentiments towards code reviews, with a majority self-reporting their value as a beneficial learning experience and agreeing that similar policies should be used in future semesters.

\edit{The university in which CS1-CR was implemented had access to undergraduate and graduate TAs for the course, alongside a well-staffed tutoring lab with trained undergraduate students prepared to conduct the code reviews for the course. We believe similar implementations of these assessments could be reasonably done and scaled with relative ease when provided with similar resources.
}

\appendix
\section{Appendix: Code Review Interview Protocol}
\label{sec:protocol}
The following is the template used to create the code review interview protocols, which are given to the TAs each week to guide their interviews.
\section*{Assn \_: (Assignment Title)}
\begin{itemize}[topsep=0pt, partopsep=0pt, itemsep=0pt, parsep=0pt]
    \item Learning objective 1
    \item Learning objective 2
    \item ...
\end{itemize}
\subsection*{Task \_ - (Task Name)}
\begin{itemize}[topsep=0pt, partopsep=0pt, itemsep=0pt, parsep=0pt]
    \item 0:00-0:30 (1/2 min) - \textbf{Warm-up} quick greeting, set purpose, locate submission
    \item 0:30-1:30 (1 min) - \textbf{Demonstrate understanding of the problem space} \textit{1 Question}
    \item 1:30-3:30 (2 min) - \textbf{Describe the solution in plain English} \textit{2 Questions}
    \item 3:30-6:30 (3 min) - \textbf{Explain your code} \textit{3 Questions}
\end{itemize}
\subsubsection*{0. Demonstrate understanding of the problem space [\_ Minutes]}
\textit{Keep the source code hidden for these questions; focus on the software plan instead}
\begin{itemize}[topsep=0pt, partopsep=0pt, itemsep=0pt, parsep=0pt]
    \item Give a brief summary of the problem you were asked to solve.
    \item What new programming features did you need to use in this task?
    \item What made testing this program difficult?
    \item \textbf{Review this task’s design document}
        \begin{itemize}[topsep=0pt, partopsep=0pt, itemsep=0pt, parsep=0pt]
            \item Is it merely a copy of the instructions?
            \item Does it contain pseudocode or actual Python code? \textit{Encourage English-like pseudocode}
            \item Can you determine whether it is a plan \textit{(forward-thinking)} or a summary \textit{(hindsight)}?
        \end{itemize}
\end{itemize}
\noindent
{
\small
\begin{tabularx}{\textwidth}{| L | L | L |}
    \hline 
     \textbf{Level 2 - Clear understanding} & \textbf{Level 1 - Partial understanding} & \textbf{Level 0 - No understanding} \\\hline
     Comprehensive design document & Design is short and lacks details & Design is missing or is a copy of assignment description \\
     Accurate one-sentence summary & Basic but incomplete summary & No clear problem statement \\
     Realistic failure case given & Weak/unclear failure case & Off-topic/vague answers \\
     Multiple important aspects noted & Mentions 1-2 aspects only & No failure cases or key aspects \\\hline
\end{tabularx}
}
\subsubsection*{1. Describe the solution in plain English [\_ Minutes]}
\textit{Keep the source code hidden for these questions; focus on the software plan instead}
\begin{itemize}[topsep=0pt, partopsep=0pt, itemsep=0pt, parsep=0pt]
    \item \textbf{If AI was used on this task, to what extent was it used and was it useful?}
    \item \textit{Specific questions related to task description}
\end{itemize}
{
\small
\noindent
\begin{tabularx}{\textwidth}{| L | L | L | L |}
    \hline
     \textbf{Level 3 - Strong explanation} & \textbf{Level 2 - Adequate explanation} & \textbf{Level 1 - Minimal explanation} & \textbf{Level 0 - No explanation} \\\hline
     Comprehensive pseudocode & Pseudocode lacks some detail & Pseudocode is short and lacks many details & Pseudocode is missing or is a copy of source code \\
     Clear, complete plain-English walk-through & Reasonably clear step-by-step description & Basic summary, missing steps or detail & Cannot explain program in plain English \\
     Highlights important features and trade-offs & Identifies key features of solution & Mentions some features but incomplete & No mention of features or approach \\
     Justifies approach logically & States why chosen approach works & Approach not clearly justified & Description unclear or off-topic \\\hline
\end{tabularx}
}
\subsubsection*{2. Explain your code [\_ Minutes]}
\textit{Run their program and review their source code while discussing these questions}
\begin{itemize}[topsep=0pt, partopsep=0pt, itemsep=0pt, parsep=0pt]
    \item \textbf{Walk me through your design doc and map each section to a code region.}
    \item Specific questions related to task requirements for code quality/program behavior
\end{itemize}
\textbf{Rules}
\begin{itemize}[topsep=0pt, partopsep=0pt, itemsep=0pt, parsep=0pt]
    \item Constraints given in task requirements
\end{itemize}
\textbf{Test Cases:} \textit{Run a set of test cases with specific inputs and expected outputs}\\
{
\small
\noindent
\begin{tabularx}{\textwidth}{| L | L | L | L | L |}
    \hline
    \textbf{Level 5 - Strong understanding} & \textbf{Level 4- Good understanding} & \textbf{Level 3 - Adequate understanding} & \textbf{Level 2 - Weak understanding} & \textbf{Level 0 - No understanding} \\\hline
    Thorough, complete walk-through of code & Clear, accurate explanation of most code passages & Reasonably clear explanation of main code & Explains some passages, missing or vague  & Cannot explain code passage(s) in plain English \\
    Precise explanations of all variables, functions, libraries & Justifies naming, structure, or loop/library choices & Can identify purpose of most variables/functions & Partial recognition of variables/functions & Gives wrong or irrelevant explanation \\
    Strong, detailed responses to assignment-specific questions & Addresses assignment-specific questions reasonably & Handles some assignment-specific questions correctly & Struggles with assignment-specific questions & Avoids/ignores assignment-specific code questions \\
    Anticipates edge cases and explains consequences of modifications & Demonstrates awareness of design trade-offs & Recognizes at least one consequence of modifying code & Cannot articulate consequences of modifying code & Unaware of consequences of modifying code \\
    & Reflects on AI use honestly and insightfully & & Cannot explain AI use & \\
    \hline
\end{tabularx}
}

\bibliographystyle{ACM-Reference-Format}
\bibliography{references}

@article{cho2011concerns,
author = {YoonJung Cho and Myoungsook Kim and Marilla D. Svinicki and Mark Lowry Decker},
title = {Exploring teaching concerns and characteristics of graduate teaching assistants},
journal = {Teaching in Higher Education},
volume = {16},
number = {3},
pages = {267--279},
year = {2011},
publisher = {Routledge},
doi = {10.1080/13562517.2010.524920},
URL = { 
    
        https://doi.org/10.1080/13562517.2010.524920
},
eprint = { 
    
        https://doi.org/10.1080/13562517.2010.524920
}}

@inproceedings{dick2005interviews,
author = {Dick, Martin},
title = {Student interviews as a tool for assessment and learning in a systems analysis and design course},
year = {2005},
isbn = {1595930248},
publisher = {Association for Computing Machinery},
address = {New York, NY, USA},
url = {https://doi.org/10.1145/1067445.1067456},
doi = {10.1145/1067445.1067456},
booktitle = {Proceedings of the 10th Annual SIGCSE Conference on Innovation and Technology in Computer Science Education},
pages = {24–28},
numpages = {5},
location = {Caparica, Portugal},
series = {ITiCSE '05}
}

@Article{mcleod2025gta,
AUTHOR = {McLeod, Claire L. and Almquist, Catherine B. and Ess, Madeline P. and Zhang, Jing and Schultz, Hannah and Nguyen, Thao and Tran, Khue and Hughes, Michael},
TITLE = {Graduate Teaching Assistants (GTAs): Roles, Perspectives, and Prioritizing GTA Workforce Development Pathways},
JOURNAL = {Education Sciences},
VOLUME = {15},
YEAR = {2025},
NUMBER = {7},
ARTICLE-NUMBER = {838},
URL = {https://www.mdpi.com/2227-7102/15/7/838},
ISSN = {2227-7102},
DOI = {10.3390/educsci15070838}
}

@inproceedings{mirza2019uta,
author = {Mirza, Diba and Conrad, Phillip T. and Lloyd, Christian and Matni, Ziad and Gatin, Arthur},
title = {Undergraduate Teaching Assistants in Computer Science: A Systematic Literature Review},
year = {2019},
isbn = {9781450361859},
publisher = {Association for Computing Machinery},
address = {New York, NY, USA},
url = {https://doi.org/10.1145/3291279.3339422},
doi = {10.1145/3291279.3339422},
booktitle = {Proceedings of the 2019 ACM Conference on International Computing Education Research},
pages = {31–40},
numpages = {10},
location = {Toronto ON, Canada},
series = {ICER '19}
}

@article{durak2019flipped,
author = {Yildiz- Durak, Hatice},
year = {2019},
month = {03},
pages = {073563311982795},
title = {Modeling Different Variables in Learning Basic Concepts of Programming in Flipped Classrooms},
volume = {58},
journal = {Journal of Educational Computing Research},
doi = {10.1177/0735633119827956}
}

@INPROCEEDINGS{chang2018flipped,
  author={Chang, Yi-Hsing and Song, An-Ching and Fang, Rong-Jyue},
  booktitle={2018 1st IEEE International Conference on Knowledge Innovation and Invention (ICKII)}, 
  title={The Study of Programming Language Learning by Applying Flipped Classroom}, 
  year={2018},
  volume={},
  number={},
  pages={286-289},
  doi={10.1109/ICKII.2018.8569171}}

@inproceedings{mohamed2020flipped,
author = {Mohamed, Abdallah},
title = {Evaluating the Effectiveness of Flipped Teaching in a Mixed-Ability CS1 Course},
year = {2020},
isbn = {9781450368742},
publisher = {Association for Computing Machinery},
address = {New York, NY, USA},
url = {https://doi.org/10.1145/3341525.3387395},
doi = {10.1145/3341525.3387395},
booktitle = {Proceedings of the 2020 ACM Conference on Innovation and Technology in Computer Science Education},
pages = {452–458},
numpages = {7},
location = {Trondheim, Norway},
series = {ITiCSE '20}
}

@inproceedings{alhazbi2016flipped,
author = {Alhazbi, Saleh},
year = {2016},
month = {12},
pages = {441-444},
title = {Using flipped classroom approach to teach computer programming},
doi = {10.1109/TALE.2016.7851837}
}

@book{sams2012flip,
  title={Flip your classroom: Reach every student in every class every day},
  author={Sams, Aaron and Bergmann, Jonathan},
  year={2012},
  publisher={International Society for Technology in Education/ISTE}
}

@article{sarawagi2014flipped,
author = {Sarawagi, Namita},
title = {A flipped CS0 classroom: applying Bloom's taxonomy to algorithmic thinking},
year = {2014},
issue_date = {June 2014},
publisher = {Consortium for Computing Sciences in Colleges},
address = {Evansville, IN, USA},
volume = {29},
number = {6},
issn = {1937-4771},
journal = {J. Comput. Sci. Coll.},
month = jun,
pages = {21–28},
numpages = {8}
}

@article{abidin2024flipped,
author = {Zainal Abidin, Noor Azlinda},
year = {2024},
month = {11},
pages = {25-44},
title = {The Efficacy of Flipped Classroom Models in Improving Student Engagement and Achievement: A Meta-Analysis},
volume = {2},
journal = {Global Synthesis in Education Journal},
doi = {10.61667/v180e591}
}

@article{cheng2018flipped,
author = {Cheng, Li and Ritzhaupt, Albert and Antonenko, Pavlo "Pasha},
year = {2018},
month = {10},
pages = {},
title = {Effects of the flipped classroom instructional strategy on students’ learning outcomes: a meta-analysis},
volume = {67},
journal = {Educational Technology Research and Development},
doi = {10.1007/s11423-018-9633-7}
}

@inproceedings{grunwald2015personalized,
author = {Grunwald, Dirk and Boese, Elizabeth and Hoenigman, Rhonda and Sayler, Andy and Stafford, Judith},
title = {Personalized Attention @ Scale: Talk Isn't Cheap, But It's Effective},
year = {2015},
isbn = {9781450329668},
publisher = {Association for Computing Machinery},
address = {New York, NY, USA},
url = {https://doi.org/10.1145/2676723.2677283},
doi = {10.1145/2676723.2677283},
booktitle = {Proceedings of the 46th ACM Technical Symposium on Computer Science Education},
pages = {610–615},
numpages = {6},
location = {Kansas City, Missouri, USA},
series = {SIGCSE '15}
}

@InProceedings{umair2026decline,
author="Umair, Muhammad Mahad
and Mukala, Patrick",
editor="Bhateja, Vikrant
and Biju, Soly Mathew
and Udgata, Siba K.",
title="Generative AI-Driven or AI-Assisted Software Code Generation and the Decline of Community Knowledge Sharing: Challenges and Future Prospects",
booktitle="Information System Design: AI and ML Applications",
year="2026",
publisher="Springer Nature Singapore",
address="Singapore",
pages="115--125",
isbn="978-981-95-0375-9"
}

@inproceedings{bernstein2025beyond,
author = {Bernstein, Seth and Rahman, Ashfin and Sharifi, Nadia and Terbish, Ariunjargal and MacNeil, Stephen},
title = {Beyond the Benefits: A Systematic Review of the Harms and Consequences of Generative AI in Computing Education},
year = {2025},
isbn = {9798400715990},
publisher = {Association for Computing Machinery},
address = {New York, NY, USA},
url = {https://doi.org/10.1145/3769994.3770036},
doi = {10.1145/3769994.3770036},
booktitle = {Proceedings of the 25th Koli Calling International Conference on Computing Education Research},
articleno = {7},
numpages = {18},
location = {
},
series = {Koli Calling '25}
}

@misc{showyourwork,
    Author = {Edwards, John},
    Title = {{JetBrains Marketplace; ShowYourWork Plugin}},
    Year = {2025},
    Url = {https://plugins.jetbrains.com/plugin/18353-showyourwork}
}

@InProceedings{rivera2020review,
author="Rivera, Victor
and Aslam, Hamna
and Naumchev, Alexandr
and de Carvalho, Daniel
and Khazeev, Mansur
and Mazzara, Manuel",
editor="Bruel, Jean-Michel
and Capozucca, Alfredo
and Mazzara, Manuel
and Meyer, Bertrand
and Naumchev, Alexandr
and Sadovykh, Andrey",
title="Towards Code Review Guideline in a Classroom",
booktitle="Frontiers in Software Engineering Education",
year="2020",
publisher="Springer International Publishing",
address="Cham",
pages="88--105",
isbn="978-3-030-57663-9"
}

@article{turner2018peer,
author = {Turner, Scott Alexander and P\'{e}rez-Qui\~{n}ones, Manuel A. and Edwards, Stephen H.},
title = {Peer Review in CS2: Conceptual Learning and High-Level Thinking},
year = {2018},
issue_date = {September 2018},
publisher = {Association for Computing Machinery},
address = {New York, NY, USA},
volume = {18},
number = {3},
url = {https://doi.org/10.1145/3152715},
doi = {10.1145/3152715},
journal = {ACM Trans. Comput. Educ.},
month = sep,
articleno = {13},
numpages = {37}
}

@article{hundhausen2009review,
author = {Hundhausen, Christopher and Agrawal, Anukrati and Fairbrother, Dana and Trevisan, Michael},
title = {Integrating pedagogical code reviews into a CS 1 course: an empirical study},
year = {2009},
issue_date = {March 2009},
publisher = {Association for Computing Machinery},
address = {New York, NY, USA},
volume = {41},
number = {1},
issn = {0097-8418},
url = {https://doi.org/10.1145/1539024.1508972},
doi = {10.1145/1539024.1508972},
journal = {SIGCSE Bull.},
month = mar,
pages = {291–295},
numpages = {5}
}

@article{indriasari2020peer,
author = {Indriasari, Theresia Devi and Luxton-Reilly, Andrew and Denny, Paul},
title = {A Review of Peer Code Review in Higher Education},
year = {2020},
issue_date = {September 2020},
publisher = {Association for Computing Machinery},
address = {New York, NY, USA},
volume = {20},
number = {3},
url = {https://doi.org/10.1145/3403935},
doi = {10.1145/3403935},
journal = {ACM Trans. Comput. Educ.},
month = sep,
articleno = {22},
numpages = {25}
}

@article{topping1998peer,
author = {Keith Topping},
title ={Peer Assessment Between Students in Colleges and Universities},
journal = {Review of Educational Research},
volume = {68},
number = {3},
pages = {249-276},
year = {1998},
doi = {10.3102/00346543068003249},
URL = {https://doi.org/10.3102/00346543068003249},
eprint = {https://doi.org/10.3102/00346543068003249}
}

@article{iannone2012oral,
    author = {Iannone, P. and Simpson, A.},
    title = {Oral assessment in mathematics: implementation and outcomes},
    journal = {Teaching Mathematics and its Applications: An International Journal of the IMA},
    volume = {31},
    number = {4},
    pages = {179-190},
    year = {2012},
    month = {10},
    issn = {0268-3679},
    doi = {10.1093/teamat/hrs012},
    url = {https://doi.org/10.1093/teamat/hrs012},
    eprint = {https://academic.oup.com/teamat/article-pdf/31/4/179/4762864/hrs012.pdf},
}

@article{huxham2012oral,
author = {Mark Huxham and Fiona Campbell and Jenny Westwood},
title = {Oral versus written assessments: a test of student performance and attitudes},
journal = {Assessment \& Evaluation in Higher Education},
volume = {37},
number = {1},
pages = {125--136},
year = {2012},
publisher = {Routledge},
doi = {10.1080/02602938.2010.515012},
URL = {https://doi.org/10.1080/02602938.2010.515012},
eprint = { https://doi.org/10.1080/02602938.2010.515012}
}

@INPROCEEDINGS{zhao2018oral,
author = "Yitong Zhao",
title = "Impact of Oral Exams on a Thermodynamics Course Performance",
booktitle = "2018 ASEE Zone IV Conference",
year = "2018",
month = "March",
address = "Boulder, Colorado",
publisher = "ASEE Conferences",
note = {https://peer.asee.org/29617},
number = {10.18260/1-2--29617}
}

@inproceedings{gharibyan2005oral,
author = {Gharibyan, Hasmik},
title = {Assessing students' knowledge: oral exams vs. written tests},
year = {2005},
isbn = {1595930248},
publisher = {Association for Computing Machinery},
address = {New York, NY, USA},
url = {https://doi.org/10.1145/1067445.1067487},
doi = {10.1145/1067445.1067487},
booktitle = {Proceedings of the 10th Annual SIGCSE Conference on Innovation and Technology in Computer Science Education},
pages = {143–147},
numpages = {5},
location = {Caparica, Portugal},
series = {ITiCSE '05}
}

@inproceedings{ohmann2025oral,
author = {Ohmann, Peter and Novak, Ed},
title = {A Multi-Institutional Assessment of Oral Exams in Software Courses},
year = {2025},
isbn = {9798400705311},
publisher = {Association for Computing Machinery},
address = {New York, NY, USA},
url = {https://doi.org/10.1145/3641554.3701848},
doi = {10.1145/3641554.3701848},
booktitle = {Proceedings of the 56th ACM Technical Symposium on Computer Science Education V. 1},
pages = {882–888},
numpages = {7},
location = {Pittsburgh, PA, USA},
series = {SIGCSETS 2025}
}

@inproceedings{ohmann2019oral,
author = {Ohmann, Peter},
title = {An Assessment of Oral Exams in Introductory CS},
year = {2019},
isbn = {9781450358903},
publisher = {Association for Computing Machinery},
address = {New York, NY, USA},
url = {https://doi.org/10.1145/3287324.3287489},
doi = {10.1145/3287324.3287489},
booktitle = {Proceedings of the 50th ACM Technical Symposium on Computer Science Education},
pages = {613–619},
numpages = {7},
location = {Minneapolis, MN, USA},
series = {SIGCSE '19}
}

@inproceedings{lau2023ban,
author = {Lau, Sam and Guo, Philip},
title = {From "Ban It Till We Understand It" to "Resistance is Futile": How University Programming Instructors Plan to Adapt as More Students Use AI Code Generation and Explanation Tools such as ChatGPT and GitHub Copilot},
year = {2023},
isbn = {9781450399760},
publisher = {Association for Computing Machinery},
address = {New York, NY, USA},
url = {https://doi.org/10.1145/3568813.3600138},
doi = {10.1145/3568813.3600138},
booktitle = {Proceedings of the 2023 ACM Conference on International Computing Education Research - Volume 1},
pages = {106–121},
numpages = {16},
location = {Chicago, IL, USA},
series = {ICER '23}
}

@article{tian2025learners,
title = {Learners' AI dependence and critical thinking: The psychological mechanism of fatigue and the social buffering role of AI literacy},
journal = {Acta Psychologica},
volume = {260},
pages = {105725},
year = {2025},
issn = {0001-6918},
doi = {https://doi.org/10.1016/j.actpsy.2025.105725},
url = {https://www.sciencedirect.com/science/article/pii/S0001691825010388},
author = {Jinrui Tian and Ronghua Zhang}
}

@Article{zou2026depend,
AUTHOR = {Zou, Huiwen and Chan, Ka Ian and Pang, Patrick and Manditereza, Blandina and Shih, Yi-Huang},
TITLE = {To Use but Not to Depend: Pedagogical Novelty and the Cognitive Brake of Ethical Awareness in Computer Science Students’ Adoption of Generative AI},
JOURNAL = {Education Sciences},
VOLUME = {16},
YEAR = {2026},
NUMBER = {2},
ARTICLE-NUMBER = {311},
URL = {https://www.mdpi.com/2227-7102/16/2/311},
ISSN = {2227-7102},
DOI = {10.3390/educsci16020311}
}

@inproceedings{becker2023programming,
author = {Becker, Brett A. and Denny, Paul and Finnie-Ansley, James and Luxton-Reilly, Andrew and Prather, James and Santos, Eddie Antonio},
title = {Programming Is Hard - Or at Least It Used to Be: Educational Opportunities and Challenges of AI Code Generation},
year = {2023},
isbn = {9781450394314},
publisher = {Association for Computing Machinery},
address = {New York, NY, USA},
url = {https://doi.org/10.1145/3545945.3569759},
doi = {10.1145/3545945.3569759},
booktitle = {Proceedings of the 54th ACM Technical Symposium on Computer Science Education V. 1},
pages = {500–506},
numpages = {7},
location = {Toronto ON, Canada},
series = {SIGCSE 2023}
}

@inproceedings{ghimire2024coding,
  title={Coding with ai: How are tools like chatgpt being used by students in foundational programming courses},
  author={Ghimire, Aashish and Edwards, John},
  booktitle={International Conference on Artificial Intelligence in Education},
  pages={259--267},
  year={2024},
  organization={Springer}
}

@inproceedings{zastudil2023generative,
  title={Generative ai in computing education: Perspectives of students and instructors},
  author={Zastudil, Cynthia and Rogalska, Magdalena and Kapp, Christine and Vaughn, Jennifer and MacNeil, Stephen},
  booktitle={2023 IEEE Frontiers in Education Conference (FIE)},
  pages={1--9},
  year={2023},
  organization={IEEE}
}

@inproceedings{vaithilingam2022expectation,
  title={Expectation vs. experience: Evaluating the usability of code generation tools powered by large language models},
  author={Vaithilingam, Priyan and Zhang, Tianyi and Glassman, Elena L},
  booktitle={Chi conference on human factors in computing systems extended abstracts},
  pages={1--7},
  year={2022}
}

@inproceedings{adeeb2025novice,
  title={How Do Novice Programmers Solve Code-Tracing Problems When ChatGPT Is Available? A Qualitative Analysis.},
  author={Adeeb, Elmira and Muldner, Kasia},
  booktitle={Proceedings of the 2025 ACM Conference on International Computing Education Research V. 1},
  pages={421--434},
  year={2025}
}

@inproceedings{shihab2025effects,
  title={The Effects of GitHub Copilot on Computing Students' Programming Effectiveness, Efficiency, and Processes in Brownfield Coding Tasks},
  author={Shihab, Md Istiak Hossain and Hundhausen, Christopher and Tariq, Ahsun and Haque, Summit and Qiao, Yunhan and Mulanda, Brian Wise},
  booktitle={Proceedings of the 2025 ACM Conference on International Computing Education Research V. 1},
  pages={407--420},
  year={2025}
}

@inproceedings{prather2024widening,
  title={{The widening gap: The benefits and harms of generative AI for novice programmers}},
  author={Prather, James and Reeves, Brent N and Leinonen, Juho and MacNeil, Stephen and Randrianasolo, Arisoa S and Becker, Brett A and Kimmel, Bailey and Wright, Jared and Briggs, Ben},
  booktitle={Proceedings of the 2024 ACM Conference on International Computing Education Research-Volume 1},
  pages={469--486},
  year={2024}
}

@inproceedings{brender2024s,
  title={Who’s helping who? when students use chatgpt to engage in practice lab sessions},
  author={Brender, J{\'e}r{\^o}me and El-Hamamsy, Laila and Mondada, Francesco and Bumbacher, Engin},
  booktitle={International Conference on Artificial Intelligence in Education},
  pages={235--249},
  year={2024},
  organization={Springer}
}

@inproceedings{hart2023accurate,
  title={Accurate estimation of time-on-task while programming},
  author={Hart, Kaden and Warren, Christopher M and Edwards, John},
  booktitle={Proceedings of the 54th ACM Technical Symposium on Computer Science Education V. 1},
  pages={708--714},
  year={2023}
}

@inproceedings{kam2025what,
author = {Kam, Matthew and Miller, Cody and Wang, Miaoxin and Tidwell, Abey and Lee, Irene A. and Malyn-Smith, Joyce and Perret, Beatriz and Tiwari, Vikram and Kenitzer, Joshua and Macvean, Andrew and Barrar, Erin},
title = {What do professional software developers need to know to succeed in an age of Artificial Intelligence?},
year = {2025},
isbn = {9798400712760},
publisher = {Association for Computing Machinery},
address = {New York, NY, USA},
url = {https://doi.org/10.1145/3696630.3727251},
doi = {10.1145/3696630.3727251},
booktitle = {Proceedings of the 33rd ACM International Conference on the Foundations of Software Engineering},
pages = {947–958},
numpages = {12},
location = {Clarion Hotel Trondheim, Trondheim, Norway},
series = {FSE Companion '25}
}

@inproceedings{prather2025beyond,
author = {Prather, James and Leinonen, Juho and Kiesler, Natalie and Gorson Benario, Jamie and Lau, Sam and MacNeil, Stephen and Norouzi, Narges and Opel, Simone and Pettit, Vee and Porter, Leo and Reeves, Brent N. and Savelka, Jaromir and Smith, David H., IV and Strickroth, Sven and Zingaro, Daniel},
title = {Beyond the Hype: A Comprehensive Review of Current Trends in Generative AI Research, Teaching Practices, and Tools},
year = {2025},
isbn = {9798400712081},
publisher = {Association for Computing Machinery},
address = {New York, NY, USA},
url = {https://doi.org/10.1145/3689187.3709614},
doi = {10.1145/3689187.3709614},
booktitle = {2024 Working Group Reports on Innovation and Technology in Computer Science Education},
pages = {300–338},
numpages = {39},
location = {Milan, Italy},
series = {ITiCSE 2024}
}

@InProceedings{ghimire2024from,
author="Ghimire, Aashish
and Edwards, John",
editor="Olney, Andrew M.
and Chounta, Irene-Angelica
and Liu, Zitao
and Santos, Olga C.
and Bittencourt, Ig Ibert",
title="From Guidelines to Governance: A Study of AI Policies in Education",
booktitle="Artificial Intelligence in Education. Posters and Late Breaking Results, Workshops and Tutorials, Industry and Innovation Tracks, Practitioners, Doctoral Consortium and Blue Sky",
year="2024",
publisher="Springer Nature Switzerland",
address="Cham",
pages="299--307",
isbn="978-3-031-64312-5"
}

@article{prather2023weird,
  title={{“It’s weird that it knows what I want”: Usability and interactions with copilot for novice programmers}},
  author={Prather, James and Reeves, Brent N and Denny, Paul and Becker, Brett A and Leinonen, Juho and Luxton-Reilly, Andrew and Powell, Garrett and Finnie-Ansley, James and Santos, Eddie Antonio},
  journal={ACM transactions on computer-human interaction},
  volume={31},
  number={1},
  pages={1--31},
  year={2023},
  publisher={ACM New York, NY}
}

@inproceedings{vadaparty2024cs1,
author = {Vadaparty, Annapurna and Zingaro, Daniel and Smith IV, David H. and Padala, Mounika and Alvarado, Christine and Gorson Benario, Jamie and Porter, Leo},
title = {CS1-LLM: Integrating LLMs into CS1 Instruction},
year = {2024},
isbn = {9798400706004},
publisher = {Association for Computing Machinery},
address = {New York, NY, USA},
url = {https://doi.org/10.1145/3649217.3653584},
doi = {10.1145/3649217.3653584},
booktitle = {Proceedings of the 2024 on Innovation and Technology in Computer Science Education V. 1},
pages = {297–303},
numpages = {7},
location = {Milan, Italy},
series = {ITiCSE 2024}
}

@misc{vadaparty2025integrating,
      title={Integrating Large Language Models and Evaluating Student Outcomes in an Introductory Computer Science Course}, 
      author={Annapurna Vadaparty and David H. Smith IV and Samvrit Srinath and Mounika Padala and Christine Alvarado and Jamie Gorson Benario and Daniel Zingaro and Leo Porter},
      year={2025},
      eprint={2510.18806},
      archivePrefix={arXiv},
      primaryClass={cs.CY},
      url={https://arxiv.org/abs/2510.18806}, 
}

@inproceedings{xue2024does,
  title={{Does ChatGPT help with introductory programming? An experiment of students using ChatGPT in CS1}},
  author={Xue, Yuankai and Chen, Hanlin and Bai, Gina R and Tairas, Robert and Huang, Yu},
  booktitle={Proceedings of the 46th International conference on software engineering: software engineering education and training},
  pages={331--341},
  year={2024}
}

@article{jovst2024impact,
  title={The impact of large language models on programming education and student learning outcomes},
  author={Jo{\u{s}}t, Gregor and Taneski, Viktor and Karakati{\u{c}}, Sa{\u{s}}o},
  journal={Applied Sciences},
  volume={14},
  number={10},
  pages={4115},
  year={2024},
  publisher={MDPI}
}

@inproceedings{ren2019what,
author = {Ren, Yanyan and Krishnamurthi, Shriram and Fisler, Kathi},
title = {What Help Do Students Seek in TA Office Hours?},
year = {2019},
isbn = {9781450361859},
publisher = {Association for Computing Machinery},
address = {New York, NY, USA},
url = {https://doi.org/10.1145/3291279.3339418},
doi = {10.1145/3291279.3339418},
booktitle = {Proceedings of the 2019 ACM Conference on International Computing Education Research},
pages = {41–49},
numpages = {9},
location = {Toronto ON, Canada},
series = {ICER '19}
}

@inproceedings{ko2025relationships,
author = {Ko, Shao-Heng and Zahn, Matthew and Stephens-Martinez, Kristin and Velasco, Yesenia and Battestilli, Lina and Heckman, Sarah},
title = {Relationships Between Computing Students' Characteristics, Help-Seeking Approaches, and Help-Seeking Behavior in Introductory Courses and Beyond},
year = {2025},
isbn = {9798400713408},
publisher = {Association for Computing Machinery},
address = {New York, NY, USA},
url = {https://doi.org/10.1145/3702652.3744214},
doi = {10.1145/3702652.3744214},
booktitle = {Proceedings of the 2025 ACM Conference on International Computing Education Research V.1},
pages = {313–326},
numpages = {14},
location = {Charlottesville, USA},
series = {ICER '25}
}

@inproceedings{doebling2021patterns,
author = {Doebling, Augie and Kazerouni, Ayaan M.},
title = {Patterns of Academic Help-Seeking in Undergraduate Computing Students},
year = {2021},
isbn = {9781450384889},
publisher = {Association for Computing Machinery},
address = {New York, NY, USA},
url = {https://doi.org/10.1145/3488042.3488052},
doi = {10.1145/3488042.3488052},
booktitle = {Proceedings of the 21st Koli Calling International Conference on Computing Education Research},
articleno = {13},
numpages = {10},
location = {Joensuu, Finland},
series = {Koli Calling '21}
}

@article{chiu2019study,
author = {Pit Ho Patrio Chiu and Paul Corrigan},
title = {A study of graduate teaching assistants’ self-efficacy in teaching: Fits and starts in the first triennium of teaching},
journal = {Cogent Education},
volume = {6},
number = {1},
pages = {1579964},
year = {2019},
publisher = {Cogent OA},
doi = {10.1080/2331186X.2019.1579964},
URL = { 
        https://doi.org/10.1080/2331186X.2019.1579964
},
eprint = { 
        https://doi.org/10.1080/2331186X.2019.1579964
}
}

@article{elena2025perspectives,
author = {Natalie Elena Westwood and Diane S. Srivastava},
title = {Perspectives of graduate student teaching assistants on value and meaning in their role},
journal = {Journal of Microbiology \&amp; Biology Education},
volume = {26},
number = {2},
pages = {e00115-25},
year = {2025},
doi = {10.1128/jmbe.00115-25},
URL = {https://journals.asm.org/doi/abs/10.1128/jmbe.00115-25},
eprint = {https://journals.asm.org/doi/pdf/10.1128/jmbe.00115-25}
}

@techreport{mit2026aicommittee,
  author      = {{MIT Ad Hoc Committee on AI Use in Teaching, Learning, and Research Training}},
  title       = {Report of {MIT}'s Ad Hoc Committee on {AI} Use in Teaching, Learning, and Research Training},
  institution = {Massachusetts Institute of Technology},
  address     = {Cambridge, MA},
  year        = {2026},
  month       = aug,
  day         = {13},
  note        = {Co-chaired by Eric Klopfer and Sam Madden},
  url         = {https://bpb-us-e1.wpmucdn.com/sites.mit.edu/dist/d/2418/files/2026/09/AI-Committee-Final-Report-Aug-13.pdf}
}

@article{prieto1999effects,
author = {Loreto R. Prieto and Steven A. Meyers},
title = {Effects of Training and Supervision on the Self-Efficacy of Psychology Graduate Teaching Assistants},
journal = {Teaching of Psychology},
volume = {26},
number = {4},
pages = {264--266},
year = {1999},
publisher = {Routledge},
doi = {10.1207/S15328023TOP260404},
URL = {
        https://doi.org/10.1207/S15328023TOP260404
},
eprint = { 
        https://doi.org/10.1207/S15328023TOP260404
}
}

@inproceedings{zaman2025graduate,
author = {Zaman, Alina and Cook, Amy and Phan, Vinhthuy and Windsor, Alistair},
title = {Graduate Computer Science TA Perspectives on In-Person Pedagogical Training: An Experience Report},
year = {2025},
isbn = {9798400715679},
publisher = {Association for Computing Machinery},
address = {New York, NY, USA},
url = {https://doi.org/10.1145/3724363.3729072},
doi = {10.1145/3724363.3729072},

booktitle = {Proceedings of the 30th ACM Conference on Innovation and Technology in Computer Science Education V. 1},
pages = {548–554},
numpages = {7},
location = {Nijmegen, Netherlands},
series = {ITiCSE 2025}
}

@inproceedings{mcdonald2023managing,
author = {McDonald, Emma and Arevalo, Gisele and Ahmed, Sadaf and Akhmetov, Ildar and Demmans Epp, Carrie},
title = {Managing TAs at Scale: Investigating the Experiences of Teaching Assistants in Introductory Computer Science},
year = {2023},
isbn = {9798400700255},
publisher = {Association for Computing Machinery},
address = {New York, NY, USA},
url = {https://doi.org/10.1145/3573051.3593384},
doi = {10.1145/3573051.3593384},
booktitle = {Proceedings of the Tenth ACM Conference on Learning @ Scale},
pages = {120–131},
numpages = {12},
location = {Copenhagen, Denmark},
series = {L@S '23}
}

\end{document}